%% file: mnras_deepak_tempelate.tex
\documentclass[fleqn,usenatbib]{mnras}

\usepackage{newtxtext,newtxmath}

\usepackage[T1]{fontenc}

\DeclareRobustCommand{\VAN}[3]{#2}
\let\VANthebibliography\thebibliography
\def\thebibliography{\DeclareRobustCommand{\VAN}[3]{##3}\VANthebibliography}

\usepackage{graphicx}	
\usepackage{amsmath}	
\usepackage[figuresright]{rotating}
\usepackage{gensymb}
\usepackage{natbib}

\title[Statistics of Chromospheric Variables]{Statistics of BY Draconis Chromospheric Variable Stars}

\author[D. Chahal et al.]{
Deepak Chahal$^{1,2}$\thanks{E-mail: deepakchahal294@gmail.com},
Richard de Grijs$^{1,2}$,
Devika Kamath$^{1,2}$
and Xiaodian Chen$^{3}$
\\
$^{1}$School of Mathematical and Physical Sciences, Macquarie University, Balaclava Road, Sydney, NSW 2109, Australia\\
$^{2}$Research Centre for Astronomy, Astrophysics and Astrophotonics, Macquarie University, Balaclava Road, Sydney, NSW 2109, Australia\\
$^{3}$CAS Key Laboratory of Optical Astronomy, National Astronomical Observatories, Chinese Academy of Sciences, Beijing 100101, China
}

\date{Accepted 2022 June 9. Received 2022 June 2; in original form 2022 April 7}

\pubyear{2022}

\begin{document}
\label{firstpage}
\pagerange{\pageref{firstpage}--\pageref{lastpage}}
\maketitle

\begin{abstract}
We present an extensive catalogue of BY Draconis (BY Dra)-type variables and their stellar parameters. BY Dra are main-sequence FGKM-type stars. They exhibit inhomogeneous starspots and bright faculae in their photospheres. These features are caused by stellar magnetic fields, which are carried along with the stellar disc through rotation and which produce gradual modulations in their light curves (LCs). Our main objective is to characterise the properties of BY Dra variables over a wide range of stellar masses, temperatures and rotation periods. A recent study categorised 84,697 BY Dra variables from Data Release 2 of the Zwicky Transient Facility based on their LCs. We have collected additional photometric data from multiple surveys and performed broad-band spectral energy distribution fits to estimate stellar parameters. We found that more than half of our sample objects are of K spectral type, covering an extensive range of stellar parameters in the low-mass regime (0.1--1.3 M$_{\odot}$). Compared with previous studies, most of the sources in our catalogue are rapid rotators, and so most of them must be young stars for which a spin-down has not yet occurred. We subdivided our catalogue based on convection zone depth and found that the photospheric activity index, $S_{\rm ph}$, is lower for higher effective temperatures, i.e., for thinner convective envelopes. We observe a broad range of photospheric magnetic activity for different spectral classes owing to the presence of stellar populations of different ages. We found a higher magnetically active fraction for K- than M-type stars.
\end{abstract}




\section{Introduction}

In the solar atmosphere, the layers of plasma located above the stellar photosphere exhibit temperature inversion. The temperature increases to $\sim 10^{4}$ K in the chromosphere, and it peaks at $\sim 10^6$ K in the corona. This heating of the solar atmosphere to such high temperatures is owing to the dissipation of energy carried by magneto-hydrodynamic waves and/or from reconnection of stressed magnetic field lines. The exact physics governing this situation is not yet fully understood. Magnetic fields generated through the solar dynamo process control the structure of the Sun's atmosphere. This leads to phenomena such as large and rapidly evolving starspots (cooler regions where convection is suppressed by magnetic field lines) in the photosphere, strong emission lines from the chromosphere, prominent flares or superflares (sudden releases of magnetic energy) from the corona, etc. \citep{deGrijs2021}. Hence, both photospheric magnetic activity (originating in the presence of starspots and bright faculae) and chromospheric magnetic activity (originating in chromospheric emission lines) are indicators of stellar magnetic activity in general.

Stellar magnetic activity is likely triggered by the interplay of rotation and turbulent convection at the stellar surface. Stellar rotation plays an important role in the dynamo efficiency and affects a star's magnetic activity. With increasing age, the loss of angular momentum induced by magnetised winds and structural variations leads to a rotational spin-down and, hence, it reduces the dynamo efficiency, in turn leading to reduced magnetic activity \citep{Parker1955,Skumanich1972}. Therefore, we expect young, rapidly rotating late-type stars to exhibit significant magnetic activity. There exist strong correlations among age, stellar activity and rotation period \citep{Hall2008,deGrijs2021}. Magnetic activity is closely linked to variations in the stellar magnetic field, and it is therefore connected to the structure of the stellar subsurface zone, stellar rotation and the regeneration of the magnetic field through self-sustaining dynamo activity.

These regenerated magnetic field lines produce inhomogeneously distributed dark spots and bright faculae in the stellar photosphere. These features move across the stellar disc through rotation, giving rise to a gradual modulation of the star's luminosity. Such stars are generally referred to as rotational modulators or rotational variables. Fitting a rotational variable's light curves (LCs) with a multiple-order Fourier function allows us to infer its rotation period, which is the key parameter required to explain the stellar activity. The rotation periods of thousands of main-sequence rotational variables have been obtained from {\sl Kepler} observations \citep{Reinhold2013,Mcquillan2014,Reinhold2020}. The amplitude of the LC fluctuations can be used as a proxy of their photospheric magnetic activity \citep{Mathur2014b,Salabert2016}. Larger LC amplitudes imply more significant magnetic activity, i.e., larger fractions or sizes of starpots on a star's surface.

A special class of rotational variables that show strong emission lines in their chromospheres are BY Draconis (BY Dra) variables. BY Dra stars are low-mass main-sequence stars which exhibit low variability amplitudes with periods of a few days; Ca {\sc ii} H and K (and often hydrogen) emission lines are usually present in their spectra \citep{Bopp1977}. The variability amplitude of BY Dra, caused by spots and faculae, can reach on the order of several tenths of a magnitude for the most active objects, but it may only be on the order of milli-magnitudes for solar-type stars. Recent observations obtained by the {\sl Kepler Space Telescope} \citep{Reinhold2013,Mcquillan2014,Reinhold2020}, the Zwicky Transient Facility \citep[ZTF;][]{Chen2020} and the Large Sky Area Multi-Object Fibre Spectroscopic Telescope (LAMOST) Medium-Resolution Survey \citep{Zhang2019} have provided large numbers of high-quality photometric (with precisions as good as 0.01 mag) and spectroscopic time-series observations of rotational modulators. \citet{Lanzafame2018} presented a catalogue of 147,535 BY Dra candidates, including their rotation periods and modulation amplitudes.

The exact mechanism which drives the stellar dynamo process---and, hence, regeneration of the magnetic field in rotational variables---is not yet properly understood. Therefore, understanding the relationship between stellar rotation and tracers of magnetic activity is important to understand the physics of the stellar dynamo. In addition, it is not yet clear either how the magnetic activity varies across a range of stellar rotation rates, masses and ages. Moreover, it is still uncertain how different the stellar dynamo process is from the solar dynamo's. Therefore, a comparison study with Sun-like active stars offers significant promise to constrain the stellar dynamo process. Statistical studies based on large data sets are still too few to be useful to gain a full understanding of the rotation--activity--age and activity--brightness--age relationships. More precise observations and estimates are required to construct a 'chromospheric Hertzsprung--Russell (HR) diagram'. The latter can be used to show where magnetically active stars are present and how activity varies across the HR diagram.

Using the large ZTF database of BY Dra candidates, we aim to address and quantify stellar magnetic activity across a wide range of stellar masses and rotation periods. Our catalogue \citep{Chen2020} contains stars of multiple spectral types, ranging from F- to M-type. We have investigated and characterised the physical properties of the BY Dra variables using a combination of LCs and photometric data from multiple surveys. Here we present an extensive catalogue of 78,954 BY Dra candidates with their stellar parameters (effective temperature, radius, luminosity, mass), rotation period and photospheric magnetic index in both the $g$ and $r$ bands. In Section \ref{section2}, we focus on the catalogue of ZTF BY Dra candidates. That section contains information about the photometric data we adopted from multiple surveys and our further analysis. In Section \ref{section3}, we discuss the broad-band spectral energy distribution (SED) fitting process we applied and the derived stellar parameters. All photometric data and the derived stellar parameters are included in Tables \ref{table1} and \ref{table2}, respectively. In Section \ref{section4}, we perform a statistical analysis of the rotation periods and the photospheric magnetic index, $S_{\rm ph}$. We discuss the relationships between multiple stellar parameters derived from our SED fitting with the $S_{\rm ph}$ index and rotation period for different spectral types. Finally, Section \ref{section5} summarises our results.

\section{Observations and Data Analysis} \label{section2}

In this section, we present the target sample used in this study.

\subsection{Target Sample}
Our initial target sample is drawn from the ZTF data. The ZTF is an optical time-domain survey undertaken with the 7-inch Oschin--Schmidt telescope at Palomar Observatory \citep{Masci2019}. It is equipped with a 47 deg$^{2}$ field of view, which can scan the entire visible northern sky in one night to median depths of $\emph{g} \sim 20.8$ mag and $\emph{r} \sim 20.6$ mag. 

We adopted the data set from \citet{Chen2020}, who used ZTF Data Release 2 (DR2), to search for and classify different types of variable stars. ZTF DR2 contains data acquired between 2018 March and 2019 June, a timespan of $\sim$470 days. Photometric data are available in the $\emph{g}$ and $\emph{r}$ bands, obtained with uniform exposure times of 30 s per snapshot observation. The search for periodic variables was done using Lomb--Scargle periodograms \citep{Lomb1976,Scargle1982}, while the classification was carried out using the DBSCAN method \citep[see][and references therein]{Chen2020} based on the distribution of periods, LC parameters and luminosities. \citet{Chen2020} classified 781,602 periodic variables into ${\delta}$ Scuti, EW- and EA-type eclipsing binaries, fundamental-mode (RRab) and first-overtone (RRc) RR Lyrae, classical and Type II Cepheids, semi-regular (SR) variables, Miras and rotational variables, including BY Draconis (BY Dra) and RS Canum Venaticorum (RS CVn) objects. 

\subsubsection{Classification of Rotational Variables}

\begin{figure}
    \centering
    \includegraphics[width=8cm]{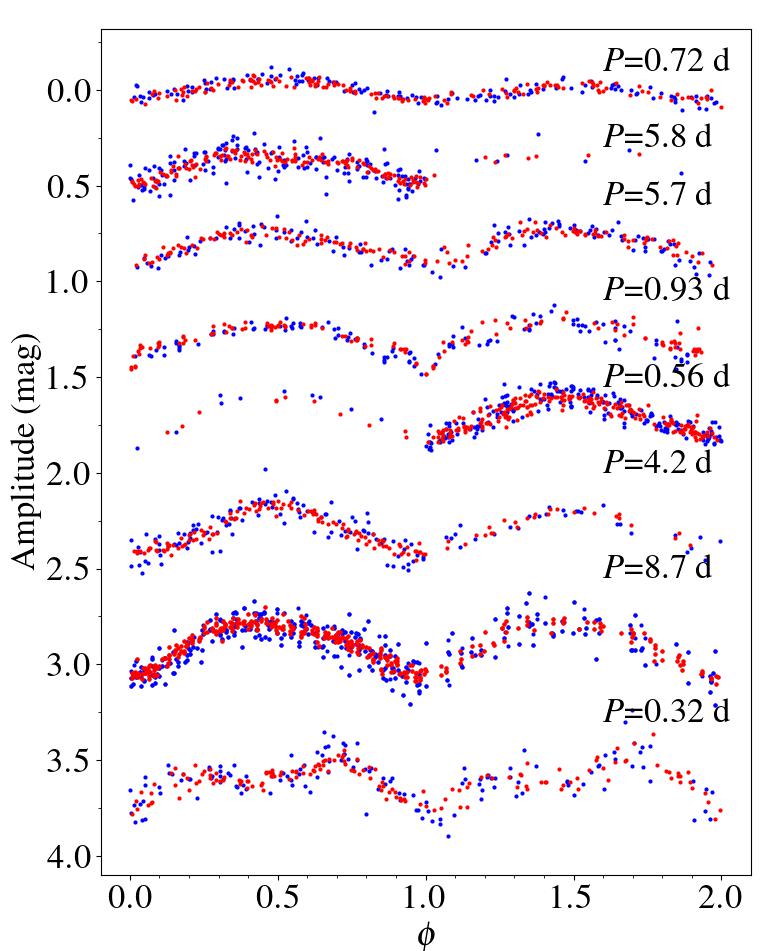}
    \caption{Example LCs of BY Dra variables. The blue and red points represent LCs in the $g$ and $r$ bands, respectively. From top to bottom, the variability amplitude increases; rotation periods are labelled for the respective LCs.}
    \label{fig:LCs}
\end{figure}

\begin{figure*}
    \centering
    \includegraphics[width=8cm]{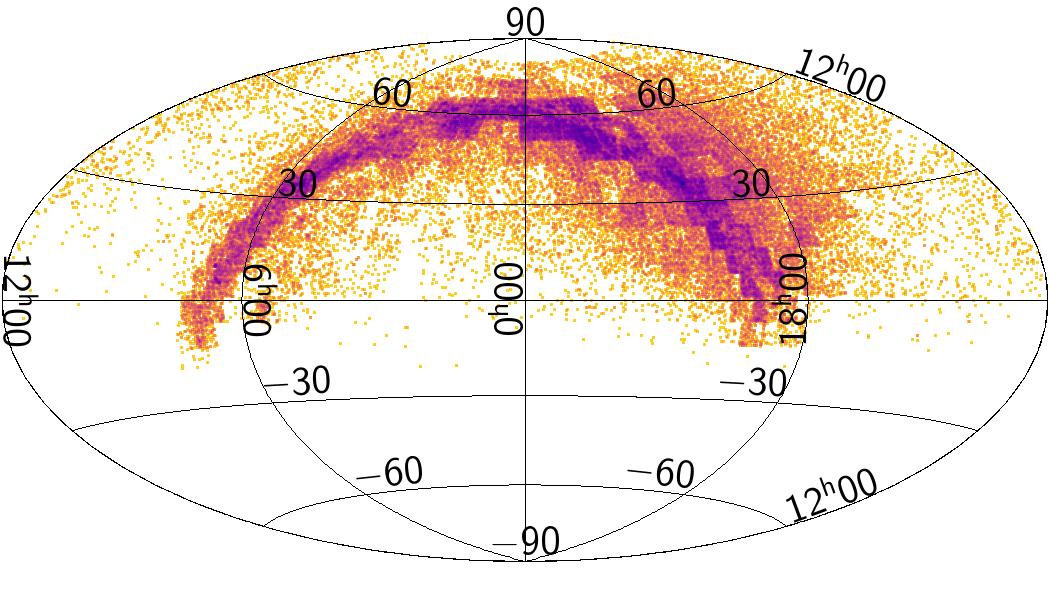}
    \includegraphics[width=9cm]{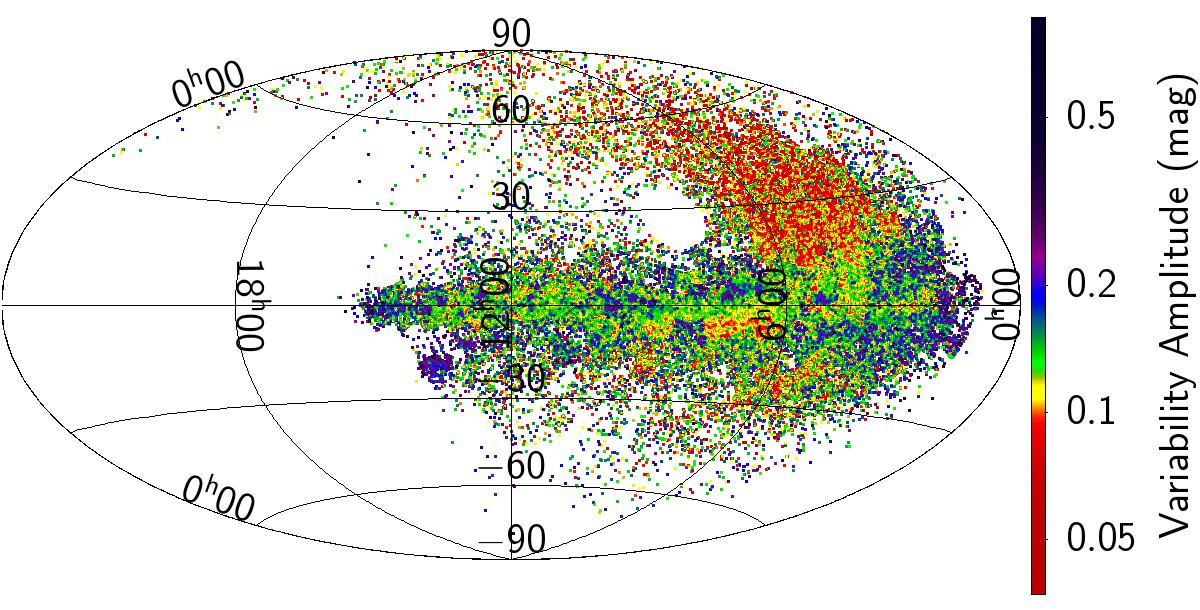}
    \caption{Distribution (left: equatorial coordinates; right: Galactic coordinates) of our sample of 78,954 ZTF BY Dra variables. The map on the right is colour-coded based on variability amplitude. }
    \label{fig:MW}
\end{figure*}

To estimate the physical parameters associated with the LCs such as the rotation periods and variability amplitudes, \citet{Chen2020} fitted the LCs with a fourth-order Fourier function. Sources with poor LC fits were excluded by optimisation of the $R^{2}$ goodness-of-fit parameter, i.e., they removed sources with $R^{2}_{g} < 0.4$ and $R^{2}_{r} < 0.4$. Rotational variables such as RS CVn and BY Dra exhibit periodic but non-characteristic LCs (see Figure \ref{fig:LCs}). BY Dra variables exhibit out-of-eclipse LC distortions arising from the rotation of spotted regions, which leads to slow changes in their mean brightness. We have found mean brightness variations in our data on the order of 0.12 mag and 0.10 mag in the \emph{g} and \emph{r} bands, respectively. A few BY Dra example LCs in the \emph{g} and \emph{r} bands are shown in Figure \ref{fig:LCs}. The LCs in both bands look similar. However, in most cases their overall shapes are not symmetric about either phase $\phi = 0.5$ or $\phi = 1.5$. These variables have small variability amplitudes compared with RR Lyrae, Cepheids or eclipsing binary systems; their rotation periods range from a fraction of a day to well over 100 days. BY Dra variables are generally late-type, main-sequence dwarf stars exhibiting enhanced magnetic activity. \citet{Chen2020} observed flaring features in several BY Dra LCs, but those data points were $\sigma$ clipped prior to LC fitting.

The classification by \citet{Chen2020} based on ZTF DR2 resulted in 84,697 and 81,293 BY Dra and RS CVn stars with purities of 94.6 per cent and 87.2 per cent, respectively. RS CVn are more evolved active stars. RS CVn and BY Dra objects exhibit similar rotation periods, although their luminosities differ significantly, with BY Dra being fainter. Our knowledge of stellar chromospheric activity will benefit greatly from a careful statistical analysis of these rotational variables. In this paper, we focus on the sample of 84,697 BY Dra variables. The full categorised ZTF catalogue is available online.\footnote{\url{http://variables.cn:88/ztf/}} Based on a careful examination of the number density distribution of periodic variable candidates, \citet{Chen2020} imposed a cut-off at a period of 20 days, which seems a natural cut-off for BY Dra variables. This cut-off also removed distant semi-regular variables which may have been misclassified as BY Dra variables, since it is difficult to distinguish between the two types of variables if {\sl Gaia} parallaxes are insufficiently accurate.

We have further refined our target sample by cross-matching our ZTF sample with the {\sl Gaia} early DR3 \citep[EDR3;][]{GaiaCollaboration2021} catalogue. \citet{Sassun2021} recently reported that a renormalised unit weight error (RUWE) goodness-of-fit statistic $\gtrsim$ 1.4 indicates likely unreliable astrometry. Objects with high RUWE values could be unresolved binary stars or tight astrometric binaries. Therefore, we have ensured that all stars in our catalogue have RUWE $< 1.4$ and are thus characterised by reliable {\sl Gaia} parallaxes \citep[as well as the corresponding Bailer-Jones geometric distances, see below;][]{BailerJones2021}). This left a target sample of 78,954 sources.

The distribution of the 78,954 BY Dra variables in the northern hemisphere is shown in Figure \ref{fig:MW}. This distribution shows that the majority of our stars are located in the Galactic disc, since ZTF has a higher cadence in the Galactic plane ($\pm$ 7$\degree$). This is different from the {\sl Kepler} sample, and so our results may indeed differ compared with work resulting from {\sl Kepler} data. Our sample also consists of stars covering a much wider age range. Moreover, the sample's distribution outside the Galactic disc is confined to the solar neighbourhood. The map in Figure \ref{fig:MW} (right) shows that the majority of stars in the solar neighbourhood have lower variability amplitudes (shown with red colours in Figure \ref{fig:MW}), possibly because of our ability to detect fainter magnitude variations at close proximity compared with more distant regimes.

\subsection{Photometric Data}

To estimate the stellar properties of our target sources we used photometric data covering the optical, near- and mid-infrared wavelength ranges. We found good cross-matches with {\sl Gaia} EDR3 \citep{GaiaCollaboration2021} for 78,954 sources within a tolerance radius of 1 arcsec. Cross-matching was done using the CDS (Centre de donn\'ees astronomiques) X-match\footnote{\url{http://cdsxmatch.u-strasbg.fr/}} tool. {\sl Gaia} EDR3 contains updated astrometry and broad-band photometry in the $G$, $G_{\rm BP}$ and $G_{\rm RP}$ bands. We have used \citet{BailerJones2021} distances---which were estimated using a probabilistic approach---for our follow-up analysis. We had already filtered out unreliable distances by removing sources with RUWE $\ge 1.4$.

To include photometric data from other surveys, we also cross-matched our sample with the infrared AllWISE catalogue \citep{Cutri2021} and the Two Micron All-Sky Survey (2MASS) All-Sky Catalog of Point Sources \citep{Cutri2003}. The 2MASS survey scanned the entire sky uniformly in three near-infrared bands ($J,H,K$). AllWISE combines data from the Wide-field Infrared Survey Explorer \citep[WISE;][]{Wright2010} and Near-Earth Objects WISE \citep[NEOWISE;][]{Mainzer2011} surveys to provide the currently most comprehensive view of the mid-infrared sky in four passbands ($W1,W2,W3,W4$). Our cross-matching with the ZTF BY Dra sample resulted in 81,479 and 64,835 sources from the 2MASS and AllWISE surveys, respectively.

The photometric uncertainties in a few of our passbands are rather large compared with the mean amplitude of variability, i.e., they exceed the mean variability amplitude of 0.1 mag in the $r$ band. We applied 2$\sigma$ clipping to remove all photometric data with uncertainties greater than 0.1 mag ($\sim$2--3 per cent of our objects). Similarly, for the parallaxes and {\sl Gaia} distances, all sources with fractional uncertainties greater than 2$\sigma$ were removed. We corrected for extinction in multiple bands using the 3D map of dust reddening of \citet{Green2019}, combined with the Galactic extinction law \citep{Wang2019}. The 2$\sigma$-clipped and extinction-corrected data are included in Table \ref{table1}, which contains source co-ordinates, {\sl Gaia} astrometry, multipassband photometry, rotation periods and variability amplitudes. The entire table is available electronically in the online journal. This photometry is used in our SED fitting, which is discussed in Section \ref{section3}.

\input{Table1}

\subsection{Colour--Magnitude Relations}

\begin{figure}
\centering
    \includegraphics[width=8.5cm]{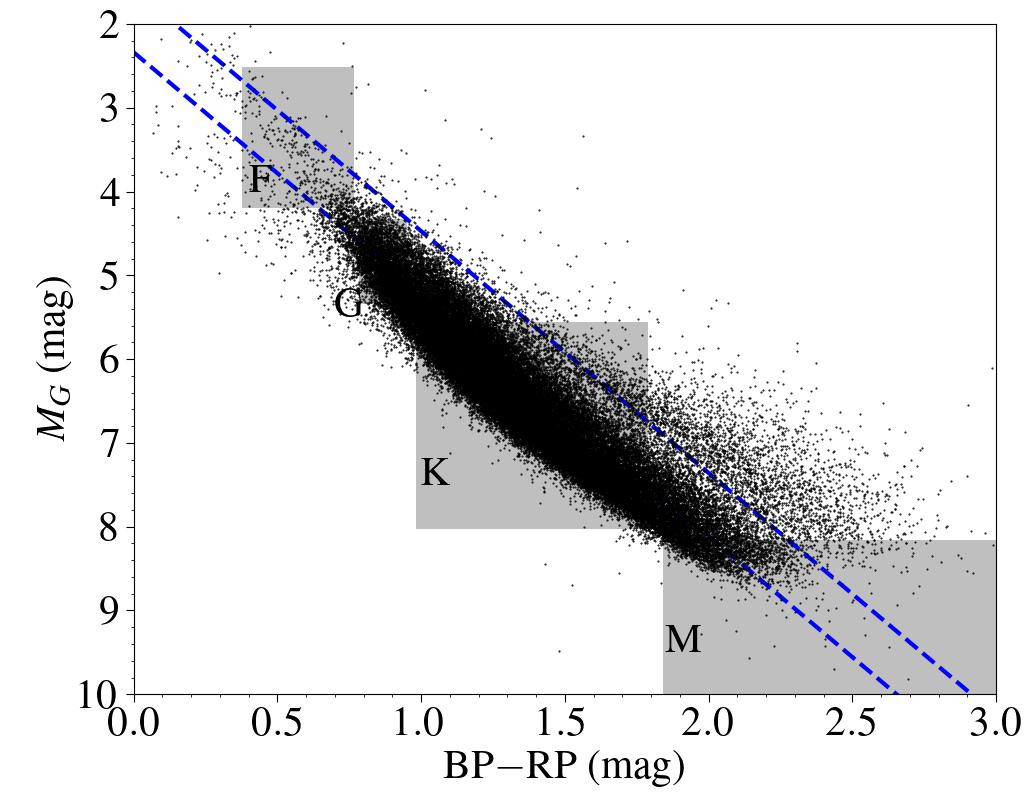}
\caption{{\sl Gaia} absolute $G$-band magnitude versus ($G_{\rm BP}-G_{\rm RP}$) diagram. Grey regions delineate the F, G, K and M spectral types \citep{Pecaut2013}. The blue dashed lines represents the means in our colour bins; equal-mass (equal-luminosity) binary systems are located 0.753 mag above the mean single-star isochrone. }
\label{fig:CMD}
\end{figure}

To be able to understand the evolutionary sequence of BY Dra variables, we show the relevant {\sl Gaia} colour--absolute magnitude diagram in Figure \ref{fig:CMD}. We compared our colours with the empirically calibrated main-sequence tables of \citet{Pecaut2013}. We found that all of our ZTF sources fall within the F-to-M spectral-type range, as shown in Figure \ref{fig:CMD}. That figure shows that most of our sources overlap with the main-sequence regime (shown as grey boxes), except for the bifurcated sources at fainter magnitudes.

To understand this bifurcation, we compared our ZTF sample with the {\sl Kepler} catalogue of rotational modulators \citep{Mcquillan2014,Reinhold2020}. Very few {\sl Kepler} sources lie along the bifurcation observed for our sample. Since previous studies have shown that most BY Dra sources also have a binary companion \citep[for a recent review, see][]{deGrijs2021}, we conclude that this bifurcation is most likely caused by the presence of binary systems in our sample, especially K- and M-type binaries. Since we do not have radial-velocity or spectroscopic data, it is beyond the scope of this paper to study the physical properties of these binaries. To classify the bifurcated sources we eliminated any nearly equal-mass binaries which are typically located 0.753 mag above the single-star main sequence (as shown by the blue dashed lines in Figure \ref{fig:CMD}).

\section{Stellar Parameter Estimates} \label{section3}

The colour--magnitude diagram offers a good handle on the stellar parameters. However, better estimates of the stellar effective temperatures can be obtained through SED fitting. In this section, we present the stellar parameters derived from our photometry (see Section \ref{section2}). We constructed the stellar SEDs and performed broad-band SED fitting. The final stellar parameters result from the specific SED model which best fits the observations, based on $\chi^2$ minimisation.

\subsection{Estimating Temperature, Luminosity and Radius based on SED Fitting} 

\begin{figure*}
\centering
    \includegraphics[width=18.5cm]{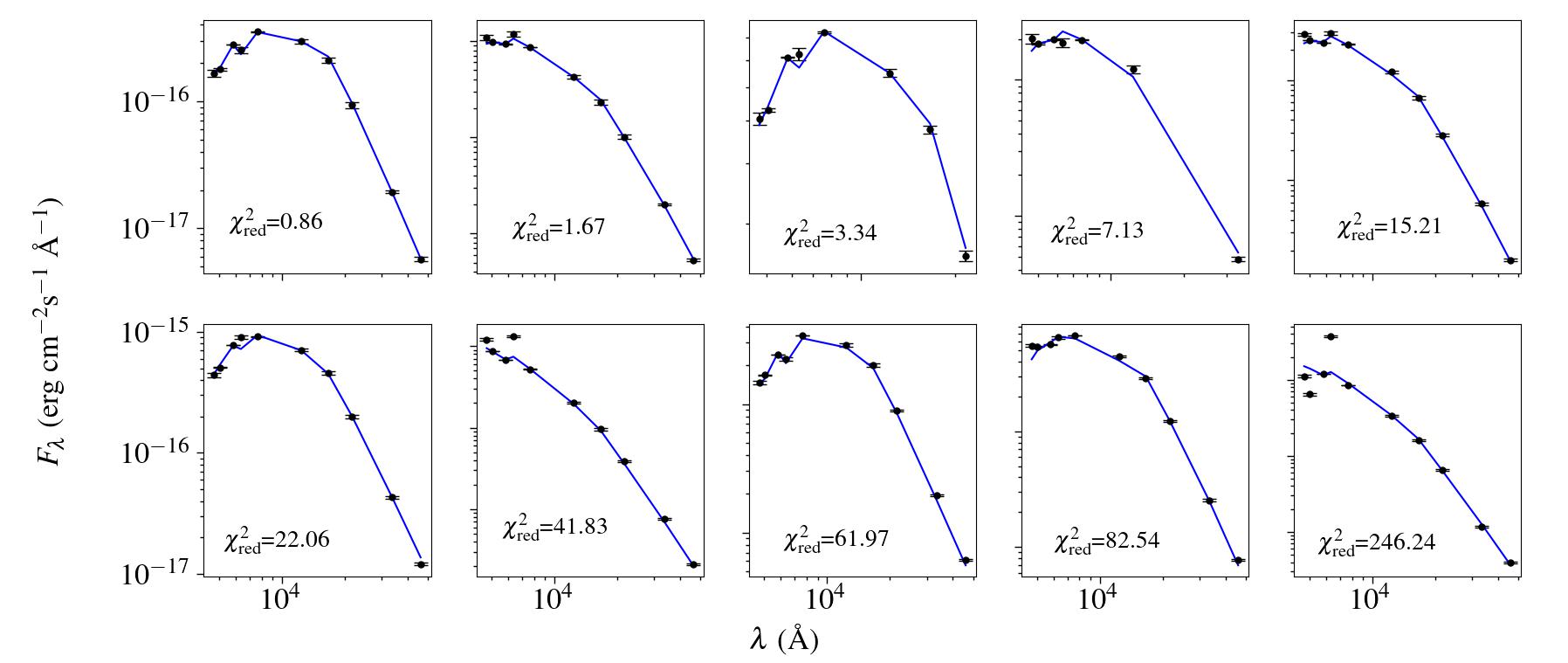}
\caption{Example SEDs of BY Dra candidates. The black filled circles with error bars are the observed photometric data points, whereas the blue line represents the best-fitting BT-Settl (CIFIST) model (see the text).}
\label{fig:SED fit}
\end{figure*}

The SEDs were constructed using the available photometric data from the ZTF, {\sl Gaia} EDR3, 2MASS and AllWISE surveys. We used the Virtual Observatory SED Analyzer\footnote{\url{http://svo2.cab.inta-csic.es/theory/vosa/}} \citep[VOSA;][]{Bayo2008} for our broad-band SED fitting. VOSA generates synthetic photometry from theoretical spectra, which are then compared with the observed data and a statistical (minimum reduced-$\chi^2$) test is performed to determine which model best reproduces the observed data. We used the BT-Settl (CIFIST) stellar atmosphere model \citep{Allard2011}. This model contains a grid of theoretical spectra, which is valid across the entire parameter range of interest here and uses the \citet{Caffau2011} solar abundances. 

The results of the broad-band SED fits are included in Table \ref{table2}. Table \ref{table2} can be accessed electronically in its entirety in the online journal. The Table \ref{table2} contains the source coordinates and stellar parameters such as $T_{\rm eff}$, $\log g$, luminosity, radius, mass, age and the photospheric magnetic index ($S_{\rm ph}$). We present a few example SED fits in Figure \ref{fig:SED fit}. That figure shows a number of best-fitting SEDs characterised by different reduced-$\chi^{2}$ values. It is evident that the sample SEDs with large reduced-$\chi^{2}$ values nevertheless fit most of the data points well. Therefore, we only removed sources for which we have fewer than seven photometric data points since such sparse wavelength coverage results in unacceptably poor SED fits. We note that many SED fits show a slight dip in the ZTF $r$ band (central wavelength $\lambda = 6250$ \AA), as can be seen in Figure \ref{fig:SED fit}. This dip might be related to the presence of an H$\alpha$ feature.

\input{Table2}

We probed $T_{\rm eff} \in [2800-7000]$ K in steps of 100 K and allowed $\log g$ to vary between 3.5 and 5.0, i.e., covering the typical range of cool and low-mass stars (including FGKM spectral types; see the colour--magnitude diagram in Section \ref{section2}) \citep{Kubiak2021}. However, note that the SED fitting procedure is mostly insensitive to $\log g$. Therefore, the main parameter constrained by our SED fitting is $T_{\rm eff}$. The luminsoities were derived from the best-fitting observed flux and the radius were estimated from the Stefan--Boltzmann equation \citep[for more details, see][]{Bayo2008}. Since the SED analysis is not sensitive to $\log g$, the resulting masses obtained from using $\log g$ are not well-determined. The $T_{\rm eff}$ values imply that more than half of our stars are of K spectral type, with the respective fractions progressively decreasing for G, M and F types (see Table \ref{table3}).

\begin{table}
\centering
\caption{Classification of BY Dra candidates based on their convection zone depth.}
\resizebox{6cm}{!}{%
\begin{tabular}{@{}ccc@{}}
\hline
\hline
Spectral Type & No. of Sources & Fraction  \\
\hline
Class I ($\leqslant$ 3900 K) & 10,336 & 12 per cent \\
Class II (3900--5400 K) & 47,119 & 56 per cent \\
Class III ($\geqslant$ 5400 K) & 6879 & 8 per cent \\ 
\hline
\end{tabular}%
}
\label{table3}
\end{table}

\begin{figure*}
\centering
    \includegraphics[width=5.5cm]{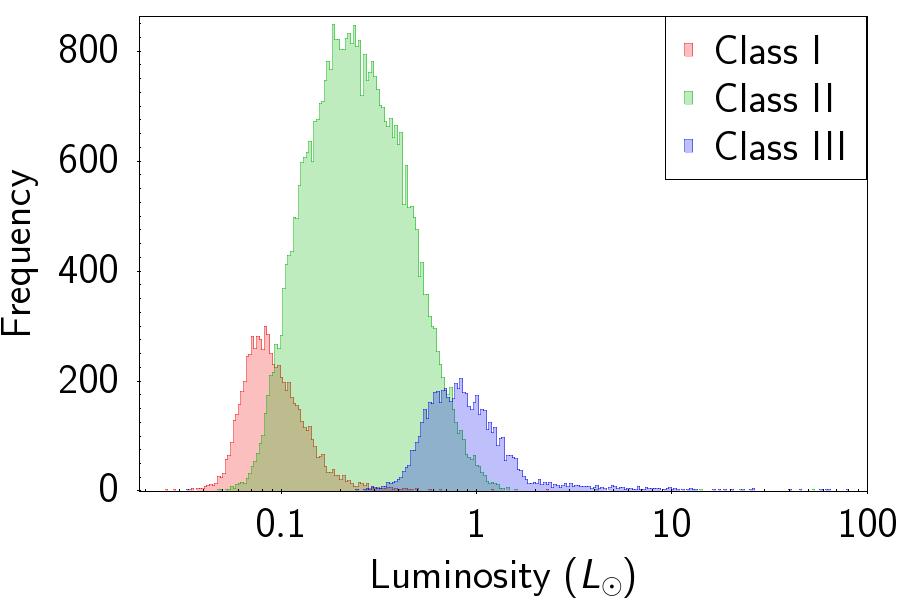}
    \includegraphics[width=5.5cm]{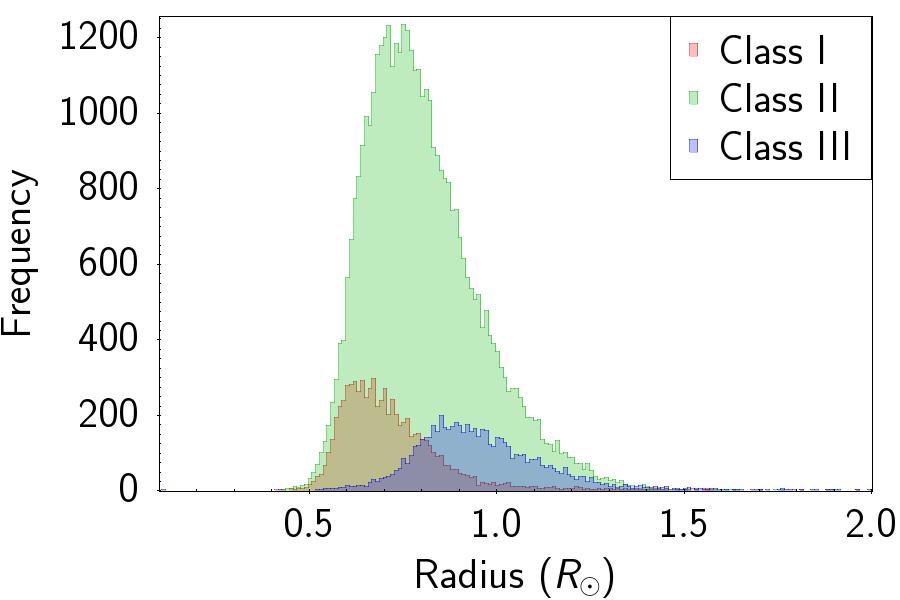}
    \includegraphics[width=5.5cm]{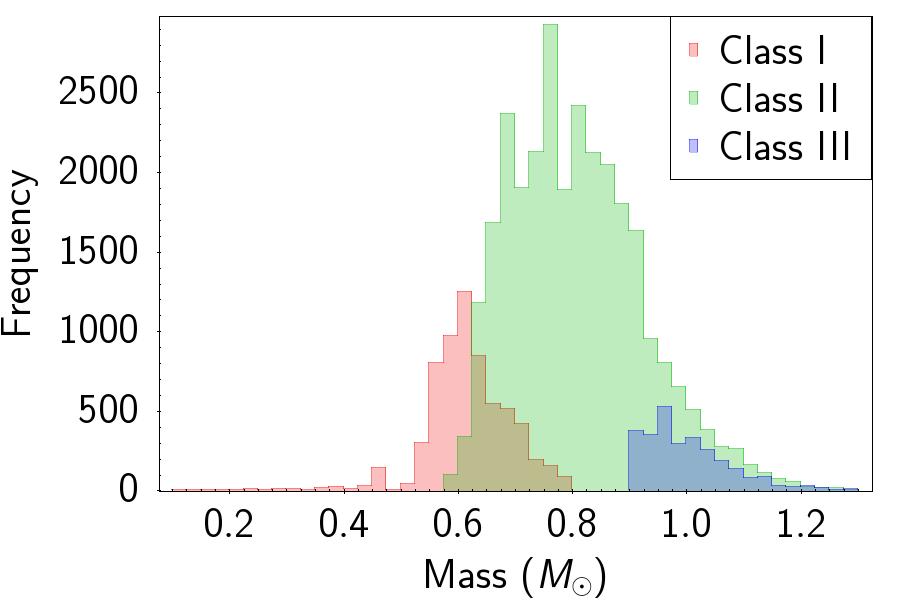}
\caption{Histogram of stellar parameters derived from the broad-band SED fits. (right) Luminosities (in units of L$_{\odot}$); (middle) radii (in units of R$_{\odot}$); (left) masses (in units of M$_{\odot}$).}
\label{fig:Histogram of Parameters}
\end{figure*}

The effective temperature, $T_{\rm eff}$, controls the width of the convective envelope in a star. As $T_{\rm eff}$ increases, the convective envelope becomes thinner. The depth of the convective envelope increases towards lower effective temperatures and masses; below $\sim 0.3$ M$_{\odot}$, stars are fully convective. The presence of a convective envelope plays a significant role in the regeneration of magnetic fields through dynamo physics \citep[][and references therein]{deGrijs2021}. Hence, stars with larger convective envelopes will be more magnetically active. Our sample, which contains predominantly K-type stars, is dominated by convection. The fraction of fainter, M-type stars is smaller. The latter are fully convective. The lower fraction of M-type stars is owing to the fact that our observations are limited by instrument capabilities. The fraction of F-type stars is just 2 per cent. This is not unexpected, since these stars have very thin outer convection zones.

To properly understand our full data set of 78,954 objects, we classified all sources into three classes based on the depths of their convective zones \citep[see][]{vanSaders2012}. Figure 1 of \citet{vanSaders2012} shows the variation of the convective zone depth with $T_{\rm eff}$. It shows steep slopes for lower and higher $T_{\rm eff}$ values and a roughly constant plateau in between. The statistics of our classification are summarised in Table \ref{table3}. Class I objects have $T_{\rm eff} \leqslant 3900$ K. Here, the depth of the convective zone increases exponentially for progressively lower effective temperatures. Class II objects have $T_{\rm eff}$ between 3900 and 5400 K, a temperature regime where the convective zone depth is approximately constant (within 5--10 per cent). Finally, Class III sources have $T_{\rm eff} \geqslant 5400$ K, where the depth of the convective zone decreases exponentially with increasing effective temperature. 

The ranges of luminosities and radii for different classes of objects are shown in Figure \ref{fig:Histogram of Parameters}. This figure shows that there is a significant overlap in radii among the different classes, whereas the luminosities are mostly distinct. Our derived radius and luminosity estimates suggest that most of our sample stars are of F to M spectral types.

\subsection{Estimating Masses and Ages from Isochrone Fitting}

With increasing age, a star spins down and, hence, its stellar activity decreases \citep{Skumanich1972}. Therefore, information related to masses and ages is important to understand the stellar activity. We estimated our stellar masses and ages by fitting a `Stellar Parameters Of Tracks with Starspots' (SPOTS) model \citep{Somers2020}. This model contains a grid of solar-metallicity stellar evolutionary tracks and isochrones which include treatment of the structural effects of starspots. The model covers the stellar mass range of 0.1--1.3 M$_{\odot}$ and a surface covering fraction from 17 to 85 per cent. In this model, stars are evolved from the pre-main-sequence to the red-giant-branch phase. 

The mean of the variability amplitude of the ZTF data is approximately 0.14 mag in the $g$ band, which results in a spot fraction of 12 per cent (assuming that the LC variability is entirely caused by starspots). Therefore, for our isochrone fitting we have adopted the model with a spot fraction of 17 per cent, i.e., the SPOT-f017 model. The estimated mass values do not have large errors, but they show clustering owing to the discrete values of $T_{\rm eff}$ contained in the model. Meanwhile, the age estimates from the isochrone fits are subject to large errors, and hence they are unreliable.

Figure \ref{fig:Histogram of Parameters} shows a histogram of the stellar masses for the different classes of objects, ranging from 0.1 to 1.3 M$_{\odot}$. The mass distributions are different for the different classes of objects, with only small overlaps. Figure \ref{fig:Histogram of Parameters} shows that our ZTF catalogue contains sources spanning a wide variety of stellar parameters in the low-mass regime, especially for Class II ($T_{\rm eff} \in [3900-5400$] K) objects.

\subsection{Rotation Periods}

\begin{figure}
\centering
   \includegraphics[width=8.5cm]{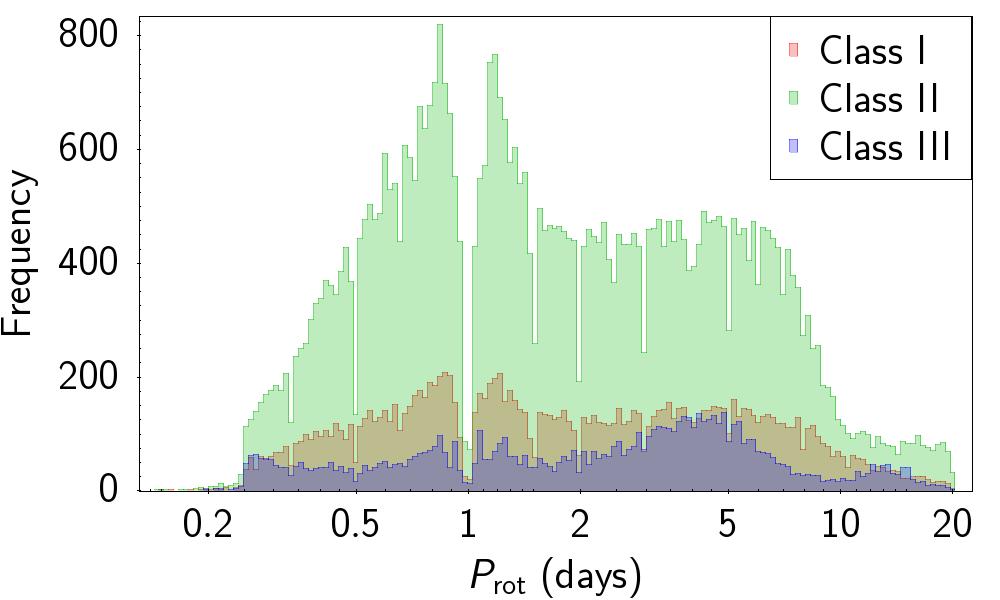}
\caption{Histogram of the rotation periods of our 65,735 BY Dra candidates, where different colours correspond to different classes of objects.
}
\label{fig:Period hist}
\end{figure}

The rotation periods of our sample objects were estimated by \citet{Chen2020}, based on fits to their LCs, as discussed in Section \ref{section2}. We have adopted the latter authors' periods. There is a possibility that some of the rotation periods in our catalogue, particularly those with values near 1 day, might be aliased periods owing to the Earth's rotation. However, \citet{Chen2020} were clearly aware of this potential bias. They attempted to avoid it from occurring as evidenced by, e.g., the distinct dearth of periods near 1 and 0.5 days evident in Figure \ref{fig:Period hist} (and also in other figures in this paper). The rotation periods of ZTF BY Dra objects range from 0.15 days to 61 days. A histogram of the rotation periods for 65,735 BY Dra candidates is shown in Figure \ref{fig:Period hist}; the mean of the distribution is 1.86 days. We found that 32 per cent, 62 per cent and 5 per cent of sources have periods of less than one day, between 1 and 10 days, and longer than 10 days, respectively. Hence, a large fraction of our sample objects rotate rapidly compared with the objects in the {\sl Kepler} catalogue of rotational modulators \citep{Mcquillan2014,Reinhold2020}. This means that a large fraction of our sample stars are young (probably in the early main-sequence phase) and, hence, the spin-down effect has not yet occurred \citep{Skumanich1972}. The statistics of the rotation periods of our ZTF BY Dra sample agree well with those pertaining to the {\sl Gaia} catalogue of rotational variables, covering late-type dwarfs \citep{Lanzafame2018}. 

In Figure \ref{fig:Period hist}, we see a clear bimodality at a rotation period of $\sim$2 days, for different classes. This bimodal period distribution at $\sim$2 days has not been observed in previous studies based on {\sl Kepler} data \citep{Mcquillan2014,Reinhold2020,Gordon2021}. The rotation distributions of the different classes show a lack of intermediate-period rotators. A similar conclusion was reached by \citet{Reinhold2020}. However, we did not find any significant decrease in the variability amplitude at $\sim$2 days. Therefore, its less likely that the bimodality arises from the non-detection of rotation periods because of the temporal increase in bright faculae, which dilutes the starspot effect \citep{Reinhold2020}.


\subsection{The Photospheric Magnetic Index}

LC variability is associated with the rotation of starspots, and thus it can be used as a proxy of stellar activity. Surface magnetic activity arises owing to the regeneration of magnetic fields through the dynamo process in the stellar interior, which is linked to the stellar rotational period. By taking into account a star's rotation period, the derived activity index, $S_{\rm ph}$, can be adopted as a proxy for its magnetic activity \citep{Mathur2014b}. The $S_{\rm ph}$ index is defined as the mean value of the standard deviation of the LC estimated over $5 P_{\rm rot}$. The measured photospheric magnetic index is only related to magnetism (i.e., starspots) and not to other sources of variability such as stellar oscillations, convective motions or instrumental issues \citep{Mathur2014b}. Since this index depends on the angle of inclination of the rotation axis relative to the line of sight, it must be considered a lower limit to the photospheric activity. \citet{Salabert2016} compared the $\langle S_{\rm ph} \rangle$ index with the spectroscopic $S$ index (the usual chromospheric activity index, derived by measuring chromospheric Ca {\sc ii} H and K line fluxes) \citep{Wilson1968} and found a good correlation.

\begin{figure}
\centering
    \includegraphics[width=8cm]{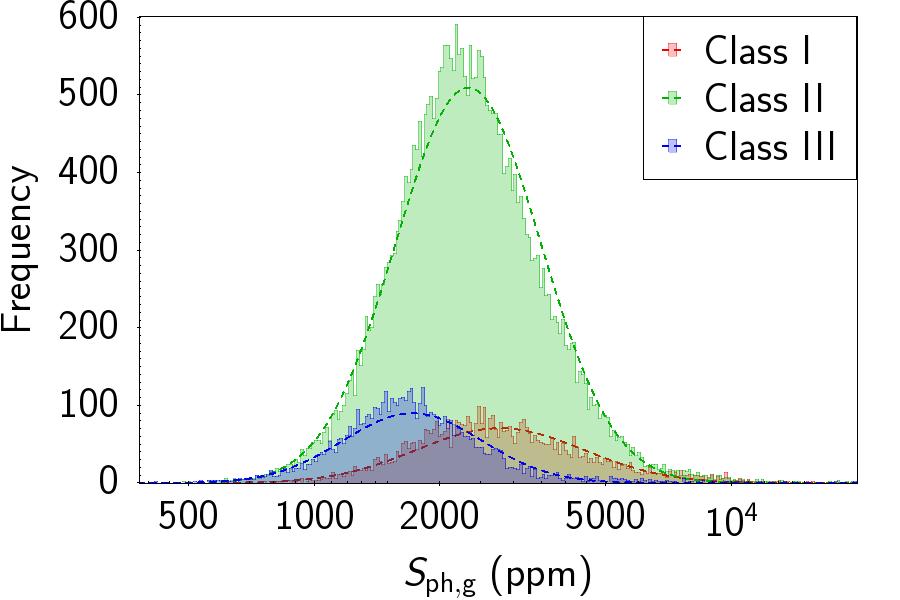}
    \includegraphics[width=8cm]{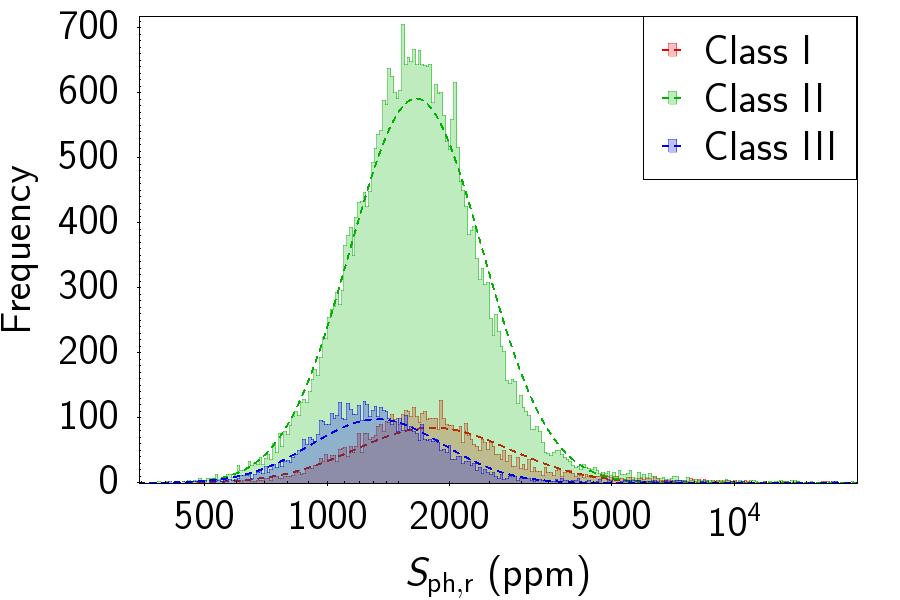}
\caption{Histograms of the photospheric magnetic index ($S_{\rm ph}$) in (top) the $g$ and (bottom) $r$ bands for different classes of objects.}
\label{fig:Sph hist}
\end{figure}

Histograms of the $S_{\rm ph}$ index in both the $g$ and $r$ bands are shown in Figure \ref{fig:Sph hist}, where different colours correspond to different classes, as specified in Section \ref{section3}. We see a clear distinction in the distributions of the different classes, although the distributions are very wide, i.e., spread over a broad range of the activity index (see Figure \ref{fig:Sph hist}). The means of the $S_{\rm ph}$ histograms in the $g$ and $r$ bands are 2486 and 1750 ppm (parts per million), respectively. Since the ZTF $r$ band is highly sensitive to chromospheric emission from active regions, the amplitude arising from the rotation of starspots is diluted by chromospheric emission in the $r$ band. 

\section{Discussion} \label{section4}

BY Dra represent a population of magnetically active K--M-type main-sequence stars which show low amplitude variability and strong chromospheric emission. Because of their low-amplitude variability, BY Dra variables have not previously been studied statistically, until the recent study by \citet{Lanzafame2018}. We are interested in understanding the stellar properties and evolutionary nature of this class of stars. These low-mass main-sequence stars exhibit low to high magnetic activity in the form of starspots, chromospheric activity, flares, etc. The observed photospheric and chromospheric activity indicators have a similar origin. However, the exact origin of these magnetic events is not yet properly understood. 

It has been argued that convection and differential rotation may play a significant role in the generation of stellar magnetic activity. Turbulent convection and differential rotation induce dynamical instabilities and, hence, play a key role in the regeneration of magnetic fields through dynamo activity. Therefore the presence of a convective envelope (which is related to stellar mass), as well as differential rotation, are the key ingredients that induce stellar magnetic activity through the dynamo process. Hence, in this section we explore the mutual dependencies of different stellar parameters such as rotation period, effective temperature, mass, etc. on stellar activity. By virtue of our large catalogue, we have managed to statistically characterise the relationships of these stellar parameters with the photospheric magnetic activity index. 



\subsection{Period--Luminosity Relation}

\begin{figure}
\centering
   \includegraphics[width=8.5cm]{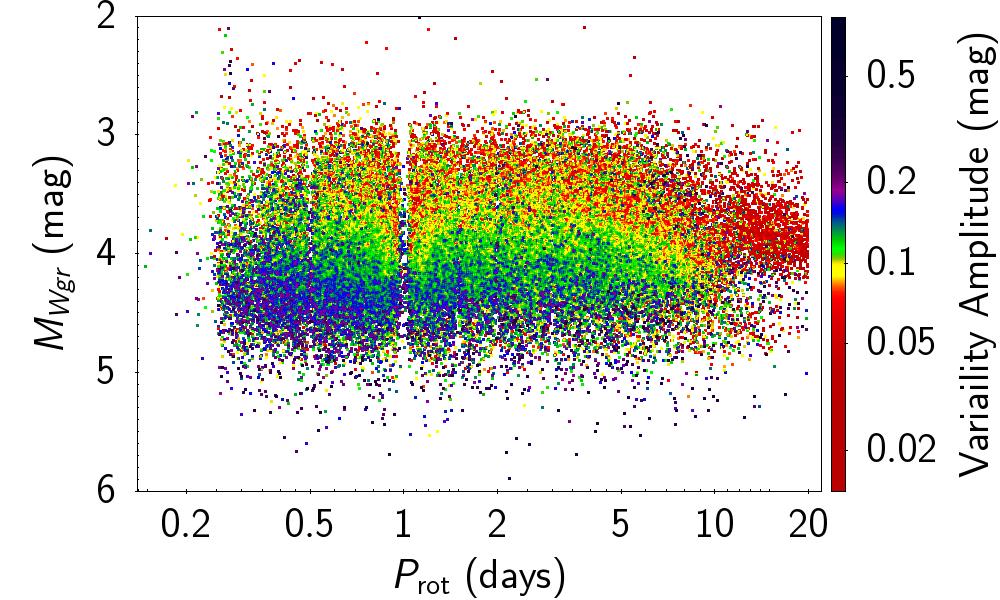}
\caption{Rotation period versus absolute Wesenheit magnitude ($M_{W_{gr}}$) diagram, colour-coded with the variability amplitude in the $\emph{g}$ band.}
\label{fig:MWgr_Per}
\end{figure}

To investigate the relationships among our photometric data, our objects' rotation periods and their variability amplitudes, we plotted the rotation periods and the ZTF absolute Wesenheit magnitudes (which are a proxy of extinction-free luminosity) in Figure \ref{fig:MWgr_Per}. The Wesenheit magnitude was obtained through $m_{W_{gr}} = 3.712\langle r \rangle - 2.712\langle g \rangle$ \citep{Chen2020}. It is evident that sources corresponding to long rotation periods (slow rotators) and high luminosities (absolute magnitudes) have lower variability amplitudes (shown in red in Figure \ref{fig:MWgr_Per}) and, hence, relatively lower fractions of starspots. Meanwhile, more rapidly rotating sources exhibit lower luminosities and, hence, higher variability amplitudes (shown in blue in Figure \ref{fig:MWgr_Per}), i.e., they have relatively larger fractions of starspots. Therefore, rapidly rotating lower-luminosity stars of late spectral types, i.e., possessing thicker convective envelopes, exhibit higher magnetic activity and, hence, larger fractions of starspots. 

\begin{figure}
\centering
   \includegraphics[width=8.5cm]{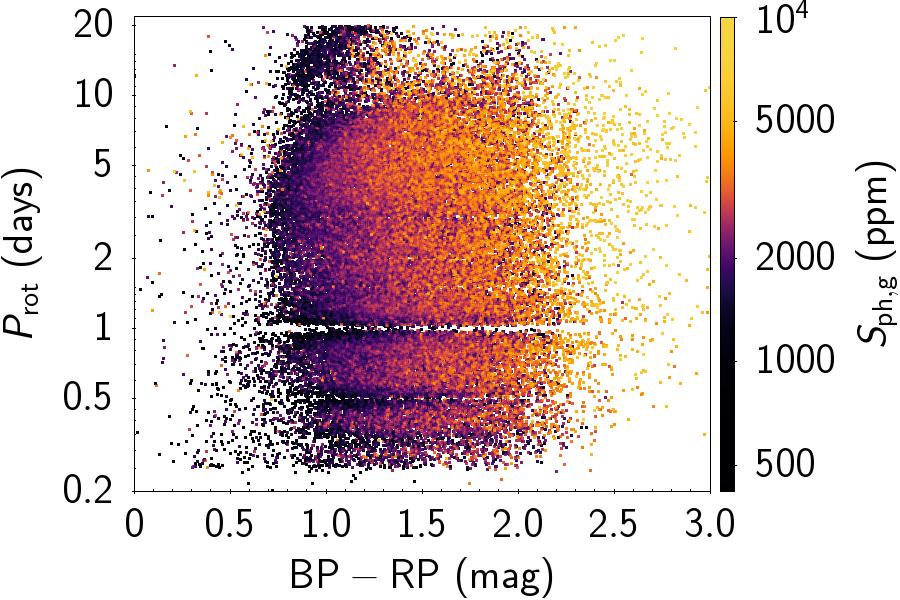}
\caption{Rotation period versus {\sl Gaia} ($G_{\rm BP}-G_{\rm RP}$) colour diagram, colour-coded with the photospheric magnetic index ($S_{\rm ph}$) in the $\emph{g}$ band.}
\label{fig:Per_Color}
\end{figure}

\begin{figure}
    \clearpage
    \centering
    \includegraphics[width=8cm]{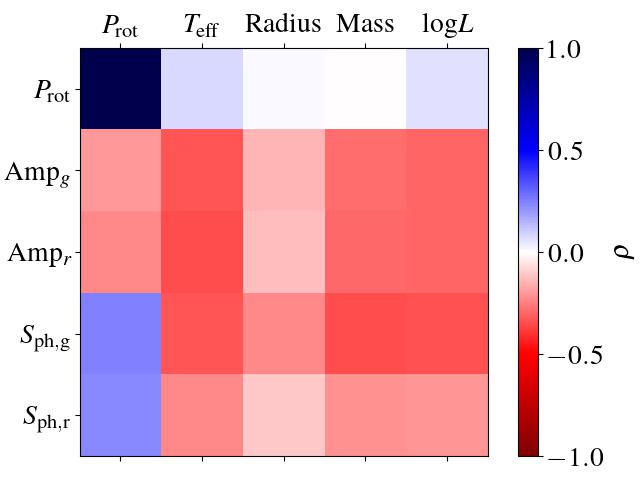}
    \caption{Spearman correlation coefficient ($\rho$) matrix of stellar parameters (rotation period, $T_{\rm eff}$, radius, luminosity and mass) and stellar activity (variability amplitude and photospheric magnetic index). }
    \label{fig:correlations}
\end{figure}

\begin{figure*}
\centering
    \includegraphics[width=8.5cm]{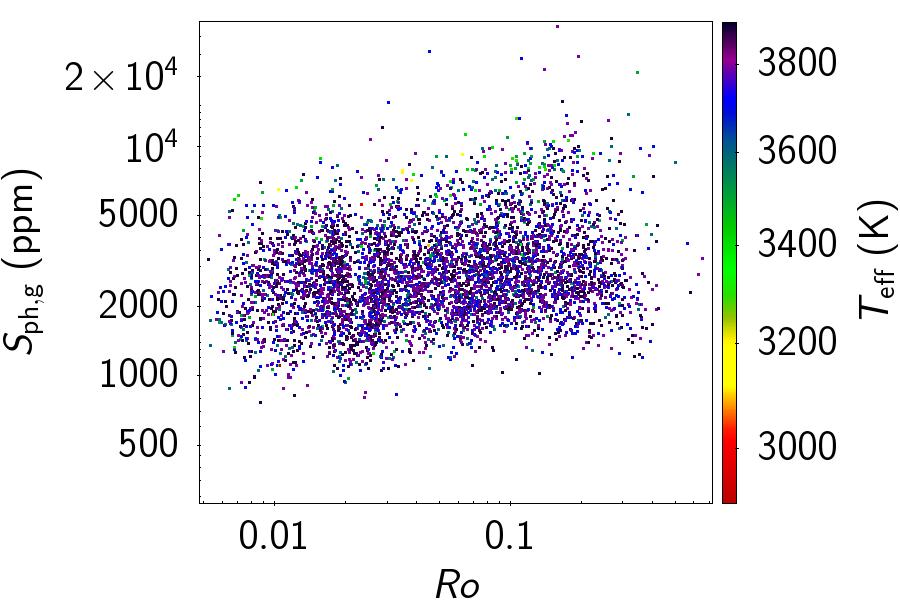}
    \includegraphics[width=8.5cm]{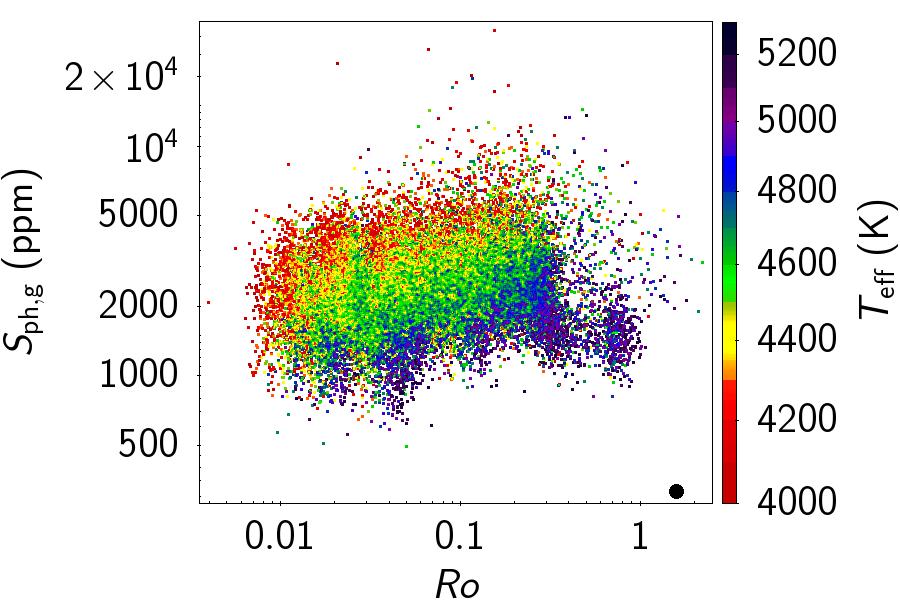}
    \includegraphics[width=8.5cm]{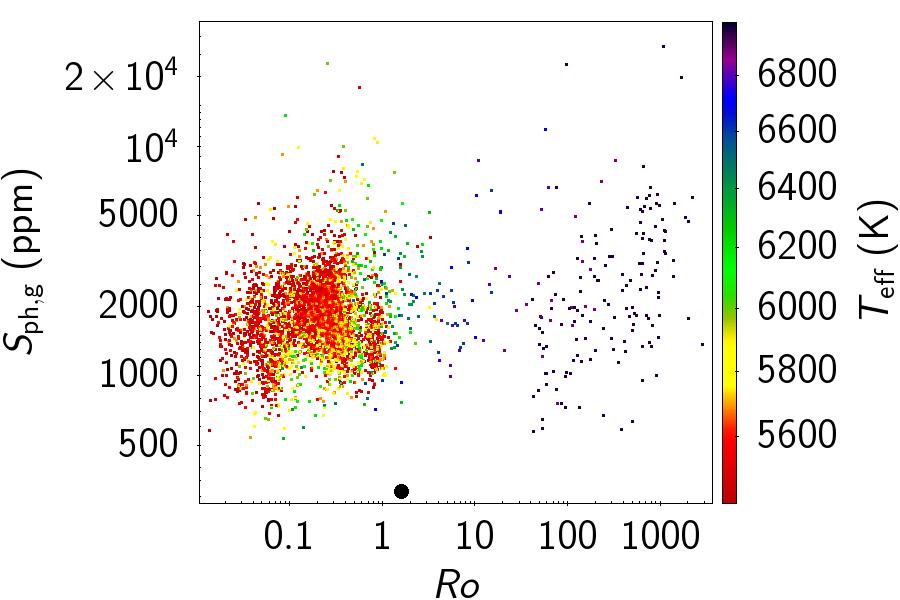}
\caption{Rossby Number versus $S_{\rm ph}$ index for different classes of objects, with the temperature distribution colour-coded. (Top left) Class I; (top right) Class II; (bottom) Class III. The black point at the bottom is the position of the Sun.}
\label{fig:Per_Sph}
\end{figure*}

\subsection{Period Gap}

As highlighted in Section \ref{section3}, we observe a bimodal distribution at a rotation period of $\sim$2 days. We also observe a bimodal distribution at the period of $\sim$10 days (see Figure \ref{fig:Period hist}). The latter bimodal distribution has been found previously in {\sl Kepler} and K2 data \citep{Mcquillan2014,Reinhold2020,Gordon2021}. In order to understand this specific bimodality in detail, in Figure \ref{fig:Per_Color} we have plotted the {\sl Gaia} colour--rotation period distribution. In this figure, we observe a small clump, and a gap below it, at a period of $\sim$10 days (see Figure \ref{fig:Per_Color}, top left). The clump and gap are also visible for a constant Rossby number ($\sim$0.5--0.7; see below) in Class II objects: see Figure \ref{fig:Per_Sph}. The gap is prominent for G--K-type stars. It appears to be a continuation of the gap previously observed in {\sl Kepler} and K2 data \citep{Mcquillan2014,Reinhold2020,Gordon2021}. \citet{Gordon2021} found that the gap extends from 15--25 days, whereas the gap in our data occurs at shorter periods and redder colours, as if it were an extension of their gap. Several explanations of the observed bimodality and associated gap have been proposed. \citet{Mcquillan2014,Davenport2018} suggested that the bimodality may be owing to a bimodal star-formation history, with an older population of stars forming a slowly rotating branch and a more recent burst of star formation corresponding to the rapidly rotating branch. Independently, \citep{Reinhold2020} suggested that these features may be owing to the non-detection of a range in rotation periods because of a temporary increase in bright faculae and, hence, dilution of the starspot effect.

\citet{Gordon2021} suggested that since the stellar core and envelope are initially not coupled, the envelope slows down because of magnetic braking, while the core may continue to spin rapidly. Magnetic braking results in a spin-down, as the star loses angular momentum through magnetic activity. After a period of time, the core and envelope start to exchange angular momentum. In turn, this counteracts the spin-down, thus resulting in an overdensity of periods below the gap. If this is indeed the case, we should have observed only a low level of activity at the bottom of the period gap, given that the spin-down would have been halted. Although we do see lower activity for F--G-type stars, we also observe a higher level of activity for K--M-type stars at the bottom of the gap (see Figure \ref{fig:Per_Color}). 

\citet{Gordon2021} also suggested that the underdensity within the gap might have been caused by an additional stage of accelerated spin-down immediately following the stalled spin-down phase. This accelerated spin-down could be owing to a temporary increase in stellar magnetic activity. If such an accelerated spin-down could have caused the observed underdensity, then the activity above the period gap should be higher relative to that in the overdense region at the bottom of the gap. However, we do not see enhanced activity. Rather, in Figure \ref{fig:Per_Color} we observe stars with a relatively low level of activity above the period gap. Since the photospheric magnetic index is prone to dilution owing to the presence of bright faculae, it tracks the minimum stellar activity. A more detailed analysis using the chromospheric activity index of the period--colour gap will offer greater constraints on the gap's origin.

\subsection{Correlation Study}

The mutual dependencies among rotation period, stellar activity and stellar parameters are complex and, hence, not yet fully understood. Since we have a large data set of magnetic variables, we have explored correlations among multiple parameters: see Figure \ref{fig:correlations}. We derived the Spearman correlation coefficients ($\rho$) for different stellar parameters and provide a correlation matrix in Figure \ref{fig:correlations}. However, we obtained low correlation coefficients for most parameter combinations because of the complex interdependencies. Still, negative correlations are discernible between certain stellar parameters ($T_{\rm eff}$, luminosity and mass) on the one hand and, on the other, the amplitude of variability and photospheric magnetic index (shown in red in Figure \ref{fig:correlations}). This is expected, since with the increase in stellar parameters (i.e., an increase in $T_{\rm eff}$ or mass), the width of the convective envelope decreases and, hence, stellar activity decreases. The stellar radii show a modest anti-correlation with stellar activity. 

We expect the stellar activity to increase for rapidly rotating stars. In Figure \ref{fig:correlations}, we see a modest negative correlation of rotation period with variability amplitude, as expected. However, we see a slight positive correlation of rotation period with $S_{\rm ph}$ index, which might be related to how the $S_{\rm ph}$ index is estimated (i.e., from the mean values of the standard deviation over $5 P_{\rm rot}$). The poor correlation of rotation period with stellar activity might be because of the relatively narrow range in rotation period covered by our sample, given that 87 per cent of our sources have rotation periods between 0.4 and 9 days. In Figure \ref{fig:correlations} we also notice the lack of clear correlations between rotation period and certain stellar parameters ($T_{\rm eff}$, luminosity, radius and mass).


\subsection{Rossby number--Activity--Temperature}

Stellar chromospheric activity correlates very well with the so-called Rossby number \citep{Noyes1984}, $Ro \equiv P_{\rm rot}/\tau_{\rm c}$, where $\tau_{\rm c}$ is the convective turnover time. The latter is estimated from the analytical equation derived from the parametrised fit as a function of effective temperature \citep[for more details, see][]{Cranmer2011}. The Rossby number probes the interplay between rotation and convection which is responsible for driving dynamo action. Figure \ref{fig:Per_Sph} shows the relationships among the Rossby number, $S_{\rm ph}$ index and $T_{\rm eff}$ for the different classes of objects. 

Among Class III objects, the width of the convective envelope becomes increasingly thinner, and therefore most of these objects are less active and, hence, have lower $S_{\rm ph}$ values. Few F-type stars (especially, $T_{\rm eff} >$ 6500 K) have large Rossby numbers because of very small values of the convective turnover time, derived from the analytical equation by \citet{Cranmer2011} (can be seen as black data-points in Figure \ref{fig:Per_Sph}). Class II objects show a clear variation with $T_{\rm eff}$; as  $T_{\rm eff}$ increases, i.e., as the convective envelope becomes thinner, we see a clear decrease in the $S_{\rm ph}$ index. However, for Class I objects, which are M-type stars, a similar trend with $T_{\rm eff}$ is not so clear, possibly owing to the increasingly thicker envelopes and eventually the absence of a tachocline, resulting in a turbulent dynamo \citep{Strassmeier2009}. The convective envelope in this class of stars is so deep that the resulting dynamo detaches from the surface or loses its connection \citep[as for tachocline stars;][]{Strassmeier2009}. 

We see a broad range in stellar activity among Classes I and II. This might be caused by the presence of a wide range of ages in the different classes. As for age, the rotation period of a star increases, i.e., the star rotates increasingly slowly owing to spin-down. Hence, stellar activity decreases with increasing age. The broad ranges in the photospheric stellar activity also arise because of the more even distribution of starspots across the stellar disc or the presence of starspots away from the line of sight, or both. Hence, we see a decrease in the variability amplitude in the LCs. Therefore, the photospheric magnetic index $S_{\rm ph}$ underestimates the magnetic activity of a star. The Sun's position is shown as the black filled circle in Figure \ref{fig:Per_Sph}. The $S_{\rm ph}$ index of the Sun was taken from \citet{Salabert2016}. The Sun belongs to a category of less active stars;  all of our sources have activity levels in excess of that of the Sun.

The distribution of Rossby number versus $S_{\rm ph}$ index shows a kink in both Classes I and II (see Figure \ref{fig:Per_Sph}). The kink corresponds to a change in the slope of the distribution from a saturated to a linear dependence of the photometric activity index on Rossby number. This kink corresponds to $Ro \approx 0.2$--0.3 and $S_{\rm ph} \approx 10^{4}$ ppm, which is in agreement with the kink's locus found by \citet{Corsaro2021} and \citet{See2021}. In coronal X-ray emission, \citet{Pizzolato2003,Reiners2014} found a similarly saturated regime at smaller Rossby numbers and subsequent linear dependence. However, we do not observe a clearly saturated region for smaller Rossby numbers as found by \citet{Corsaro2021}. Instead, we see a slight decline of the $S_{\rm ph}$ index for smaller Rossby numbers. \citet{Corsaro2021} and \citet{See2021} explored the flat, saturated region to $Ro=0.1$, whereas our data reach $Ro=0.01$. Therefore, we can say that there is slight decline in the saturation limit for very small Rossby numbers. The slope of the decline is steeper for Class I objects, which are highly active, and decreases for the other classes. This decline might be caused by bright faculae partially cancelling out dark spots, resulting in a reduction in the overall variability amplitude. Alternatively, this trend might have originated from the way the $S_{\rm ph}$ index is estimated (i.e., from the mean values of the standard deviation over $5 P_{\rm rot}$), because we did not find any such decline in variability amplitude for lower Rossby numbers. 

We also found that all categories of stars (i.e., slow, intermediate and rapid rotators) exhibit a wide range of stellar activity. To fully understand the context, age is an important parameter that must be taken into consideration. With increasing age, stars lose angular momentum and spin down, which leads to a decrease in stellar activity. Age estimates from isochrone fits have large associated errors. We will be able to constrain rotation periods, stellar parameters and magnetic activity more optimally once accurate age estimates have been obtained. We will explore ages in a future paper. 
\subsection{Chromospheric Hertzsprung--Russell diagrams}

We have some evidence that stellar magnetic activity is generated through a process akin to that governing the solar dynamo. Yet, it is still questionable whether or not the stellar dynamo which generates stellar magnetic activity is similar to that operating in the Sun. Therefore, studying stellar magnetic activity with reference to solar activity could help in understanding how different the stellar dynamo process may be from the solar dynamo for different classes of stars. 

\begin{figure}
    \clearpage
    \centering
    \includegraphics[width=8.5cm]{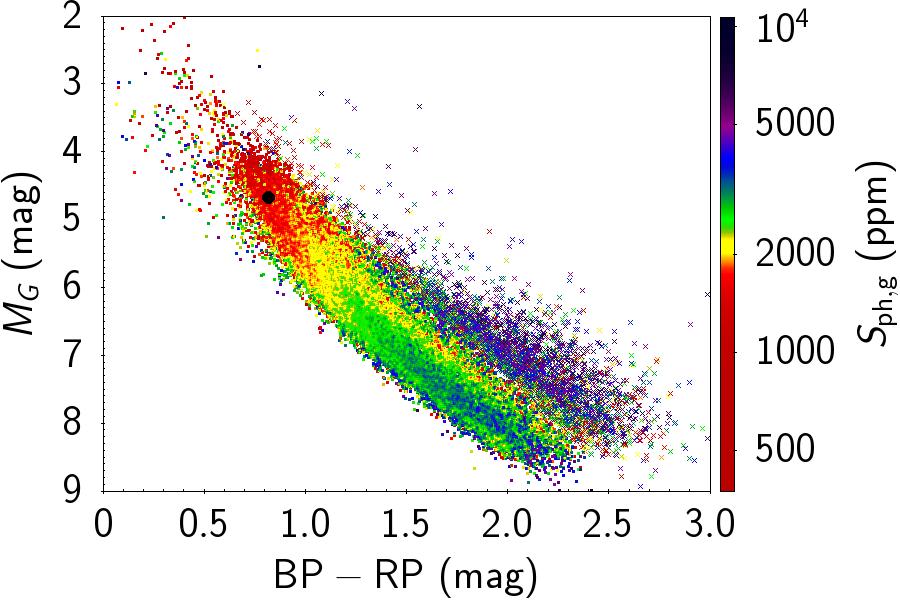}
    \caption{{\sl Gaia} absolute colour--magnitude diagram (CMD), colour-coded with the photospheric magnetic index ($S_{\rm ph}$) in the $\emph{g}$ band. The black point is the position of the Sun in the {\sl Gaia} CMD \citep{Casagrande2018}. Cross sign are the probable binary systems classified in Section \ref{section2}}
    \label{fig:Chrom HR}
\end{figure}

We show the {\sl Gaia} colour--absolute magnitude diagram of BY Dra candidates, including the density distribution of the $S_{\rm ph}$ index, in Figure \ref{fig:Chrom HR}. It shows that as the temperature increases, the convective envelope becomes thinner and, hence, the $S_{\rm ph}$ index decreases. \citet{Salabert2016} estimated the Sun's maximum $S_{\rm ph}$ index at 314.5 ppm, which is well below the $S_{\rm ph}$ index measured for the majority of the stars in our catalogue. As we have already noted, our catalogue contains very young stars (since most are rapidly rotating), and therefore we observe higher activity levels for our sample stars compared with the Sun. The Sun belongs to a class of low-to-moderately magnetically active stars. This result is also in agreement with \citet{Reinhold2020_Sun}, who compared 369 solar-like stars with the Sun.

Using our large catalogue of chromospherically active variables we constructed a `chromospheric HR diagram', which shows where Sun-like chromospheres are expected to be found. Figure \ref{fig:Chrom HR} shows that F- and late-G-type stars exhibit very low levels of stellar activity and that the activity increases towards K--M-type stars. By virtue of our large database, we conclude that most of the rotationally and magnetically active main-sequence stars in our sample are of later spectral type than F6V. We also found that late-K-type stars are most magnetically active, even more so than M-type stars. The cross signs in Figure \ref{fig:Chrom HR} show that the probable binary systems are highly active stars (visible in blue above the main sequence region at the bottom of Figure \ref{fig:Chrom HR}), which is expected since the presence of a companion increases the magnetic activity in the primary star \citep{deGrijs2021}. 

\subsection{The Putative Vaughan--Preston Gap}

\begin{figure}
\centering
    \includegraphics[width=8.5cm]{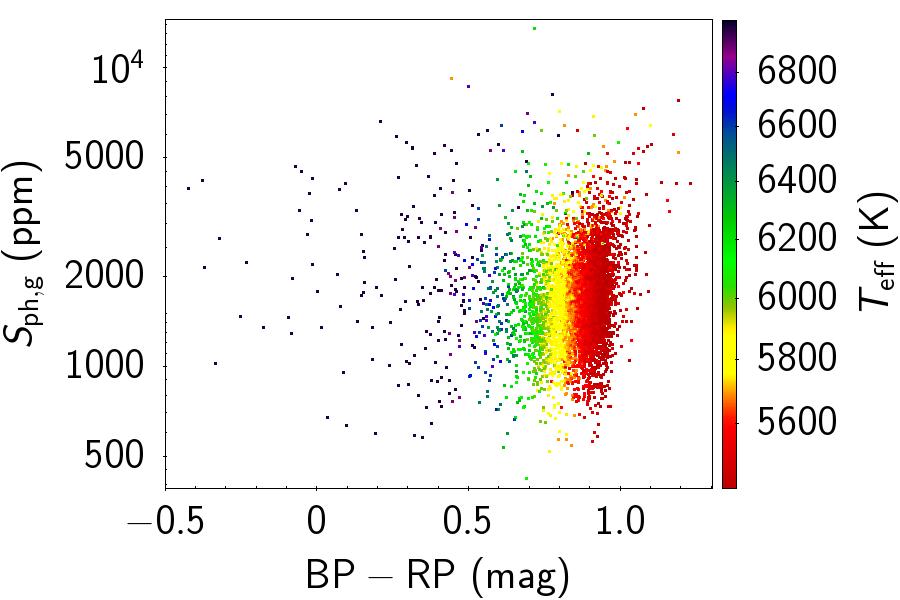}
\caption{{\sl Gaia} ($G_{\rm BP}-G_{\rm RP}$) colour versus photospheric magnetic index diagram for F- and G-type stars, with the temperature distribution colour-coded.}
\label{fig:Vaughan}
\end{figure}

An apparent dearth of F- and G-type stars of intermediate activity was discovered by \citet{Vaughan1980} based on a sample of 486 stars. We also investigated this putative `Vaughan--Preston gap' for our sample of active F- and G-type stars. We present the $S_{\rm ph}$ index versus {\sl Gaia} colour diagram in Figure \ref{fig:Vaughan} to investigate the dearth of intermediate-activity stars. Figure \ref{fig:Vaughan} shows clear signatures of the presence of stars of intermediate activity, and we observe a good spread in the magnetic activity (500--5000 ppm) of F- and G-type stars. We do not even observe a bimodal distribution in the photospheric activity index, as was seen by \citet{Saikia2018}. By virtue of our large data set, we conclude that the putative Vaughan--Preston gap may have been caused by a lack of data from stars of intermediate activity and, hence, this gap might not be a true manifestation of stellar dynamo activity.

\section{Conclusions} \label{section5}

In this paper, we have compiled a catalogue of low-mass main-sequence stars which exhibit varying levels of stellar magnetic activity. We present a large catalogue of 78,954 BY Dra candidates, including their photometric measurements, rotation periods, variability amplitudes, etc. This large catalogue has aided us in understanding the evolutionary nature of BY Dra variables. Through broad-band SED fitting, we estimated their stellar parameters, including the effective temperatures, radii, luminosities, masses, etc. We found that more than half of our sample objects are K-type stars, and the respective fraction decreases for G (12 per cent), M (8 per cent) and F (2 per cent) types. The ZTF catalogue of BY Dra candidates contains an extensive range of stellar parameters in the low-mass regime (0.1--1.3 M$_{\odot}$). 

Our ZTF catalogue contains rapidly rotating stars (94 per cent have rotation periods $\leqslant$ 10 days) compared with the {\sl Kepler} catalogue of rotational modulators \citep[see][]{Reinhold2020,Mcquillan2014}. This high rotation rate might have resulted because most of our sample stars are young (probably in the early phases of the main sequence) and, hence, the inevitable spin-down might not yet have occurred \citep{Skumanich1972}. To investigate the dependence of stellar magnetic activity, we have studied their correlation with multiple stellar parameters. We observed a negative correlation between the stellar activity (variability amplitude and $S_{\rm ph}$ index) and a number of stellar parameters ($T_{\rm eff}$, luminosity and mass), i.e., the activity decreases for hotter stars. This is expected, since with increasing $T_{\rm eff}$ or mass, the convective envelope becomes thinner, hence stellar activity decreases. We also noted a negative correlation between variability amplitude and rotation period, which is expected as rotation induces instabilities in the convection zone for the regeneration of magnetic fields through the stellar dynamo. 

To understand the different types of stars we classified our catalogue into three classes based on the slope of the convection zone depth. Class III objects ($\geqslant$5400 K) have very thin convective envelopes and, hence, yield low values of the $S_{\rm ph}$ index. Class II (3900--5400 K) objects show clear signatures of a decrease in the $S_{\rm ph}$ index as $T_{\rm eff}$ increases, i.e., the convective envelope becomes thinner. However, the trend breaks down for Class I ($\leqslant$3900 K) objects, which show a broad range of magnetic activity. This is possibly owing to the absence of a tachocline and, hence, a turbulent dynamo. The extensive range of stellar activity among all classes might be owing to the presence of a range of stellar ages for all classes. We observed a declining saturated regime in both Class I and II objects at lower Rossby numbers, and the kink is the inflexion point between the saturated and unsaturated regimes. The kink locus we observed is in agreement with that found by \citet{Corsaro2021,See2021}. Instead of a flat, saturated region for lower Rossby numbers (down to $Ro=0.01$), we observed a minor decline of the saturation limit. This might be caused by bright faculae partially cancelling out dark spots and, hence, reducing the corresponding activity index. 

We constructed a chromospheric HR diagram, which tells us that stellar activity decreases for stars hotter than early F-type, owing to the lack of a convective envelope. The activity increases towards K--M-type stars. We found that K-type stars are more magnetically active even than M-type stars. The probable binary systems in our catalogue are the most magnetically active stars. The Sun belongs to a class of low-to-moderately magnetically active stars, and the majority of our sample stars have an $S_{\rm ph}$ index greater than the Sun's. We have not found tightened relationships among stellar magnetic activity, rotation period and/or a number of other stellar parameters. Instead, we have found weak correlations among them, likely because of complex interdependencies. To fully understand these interdependencies, stellar age must be taken into consideration, since it is a key parameter that shows strong connections with both activity and rotation period.

\section*{DATA AVAILABILITY}
The ZTF data analysed in this paper are publicly available from the \citet{Chen2020} and also from the ZTF Archive. The data underlying this article are available in the article and in its online supplementary material.

\section*{Acknowledgements}
D.C. acknowledges funding support from the International Macquarie Research Excellence Scheme (iMQRES). D.K. acknowledges support from the Australian Research Council (ARC) through DECRA grant number DE190100813. This research was also supported in part by the ARC Centre of Excellence for All Sky Astrophysics in 3 Dimensions (ASTRO 3D), through project number CE170100013. X.C. acknowledges funding support from the National Natural Science Foundation of China (NSFC) through grants 12173047 and 11903045. This publication is based on observations obtained with the Samuel Oschin 48-inch Telescope at Palomar Observatory as part of the Zwicky Transient Facility project. ZTF is supported by the U.S. National Science Foundation through grant AST-1440341 and a collaboration including Caltech, IPAC, the Weizmann Institute for Science, the Oskar Klein Center at Stockholm University, the University of Maryland, the University of Washington, Deutsches Elektronen-Synchrotron and Humboldt University, Los Alamos National Laboratories, the TANGO Consortium of Taiwan, the University of Wisconsin at Milwaukee and Lawrence Berkeley National Laboratories. Operations are conducted by Caltech Optical Observatories, IPAC and the University of Washington.




\bibliography{mnemonic,example}
\bibliographystyle{aasjournal}

\bsp	
\label{lastpage}
\end{document}

%% file: Table1.tex
\begin{table*}
\caption{LC data from \citet{Chen2020} combined with photometric data of BY Dra variable candidates from multiple surveys.}
\centering
\setlength{\tabcolsep}{1pt}
\resizebox{1\textwidth}{!}{%
\begin{tabular}{@{}cccccccccccccccccccc@{}}
\hline
\hline
ID & \begin{tabular}[c]{@{}c@{}} RA (J2000)\\ (deg)\end{tabular} & \begin{tabular}[c]{@{}c@{}}Dec (J2000)\\ (deg)\end{tabular} & \begin{tabular}[c]{@{}c@{}} Parallax\\ (mas)\end{tabular} & \begin{tabular}[c]{@{}c@{}}Distance\\ (pc)\end{tabular} & \begin{tabular}[c]{@{}c@{}}$\langle g \rangle$\\ (mag)\end{tabular} & \begin{tabular}[c]{@{}c@{}}$\langle r \rangle$\\ (mag)\end{tabular} & \begin{tabular}[c]{@{}c@{}}$\langle G \rangle$\\ (mag)\end{tabular} & \begin{tabular}[c]{@{}c@{}}$\langle BP \rangle$\\ (mag)\end{tabular} & \begin{tabular}[c]{@{}c@{}}$\langle RP \rangle$\\ (mag)\end{tabular} & \begin{tabular}[c]{@{}c@{}}$\langle J \rangle$\\ (mag)\end{tabular} & \begin{tabular}[c]{@{}c@{}}$\langle H \rangle$\\ (mag)\end{tabular} & \begin{tabular}[c]{@{}c@{}}$\langle K \rangle$\\ (mag)\end{tabular} & \begin{tabular}[c]{@{}c@{}}$\langle W1 \rangle$\\ (mag)\end{tabular} & \begin{tabular}[c]{@{}c@{}}$\langle W2 \rangle$\\ (mag)\end{tabular} & ... & \begin{tabular}[c]{@{}c@{}}Period\\ (days)\end{tabular} & \begin{tabular}[c]{@{}c@{}} Amp$_{g}$\\ (mag)\end{tabular} & \begin{tabular}[c]{@{}c@{}} Amp$_{r}$\\ (mag)\end{tabular} & ... \\ \hline
ZTFJ000000.13+620605.8 & 0.00056 & 62.10163 & 1.58 & 636.77 & 16.79 & 15.73 & 15.77 & 16.56 & 14.95 & 13.83 & 13.24 & 13.06 & --- & --- & ... & 1.945 & 0.113 & 0.078 & ... \\
ZTFJ000002.20+480720.8 & 0.00918 & 48.12246 & 1.96 & 510.07 & 16.87 & 15.81 & 15.88 & 16.60 & 15.06 & 14.04 & 13.41 & 13.25 & 13.18 & 13.19 & ... & 0.381 & 0.107 & 0.093 & ... \\
ZTFJ000003.23+543605.4 & 0.01347 & 54.60151 & 1.15 & 846.92 & 17.20 & 16.22 & 16.33 & 17.03 & 15.52 & 14.45 & 13.73 & 13.73 & 13.56 & 13.63 & ... & 0.786 & 0.166 & 0.112 & ... \\
ZTFJ000003.76+532917.1 & 0.01568 & 53.48811 & 0.89 & 1116.84 & 15.84 & 15.32 & 15.35 & 15.73 & 14.79 & 14.20 & 13.74 & 13.59 & 13.66 & 13.66 & ... & 0.592 & 0.164 & 0.159 & ... \\
ZTFJ000004.34+522031.5 & 0.01809 & 52.34209 & 0.92 & 1078.8 & 17.04 & 16.28 & 16.41 & 16.94 & 15.73 & 14.98 & 14.32 & 14.26 & 14.16 & 14.22 & ... & 4.083 & 0.087 & 0.095 & ... \\
... & ... & ... &  &  & ... & ... & ... & ... & ... & ... & ... & ... & ... & ... & ... & ... & ... & ... & ...  \\
... & ... & ... &  &  & ... & ... & ... & ... & ... & ... & ... & ... & ... & ... & ... & ... & ... & ... & ... \\ \hline
\end{tabular}%
}
\label{table1}
\end{table*}

%% file: Table2.tex
\begin{table*}
\caption{Stellar Parameters of our sample BY Dra variable candidates}
\centering
\setlength{\tabcolsep}{1pt}
\resizebox{\textwidth}{!}{%
\begin{tabular}{@{}ccccccccccccccccc@{}}
\hline
\hline
$\rm {ID}$ & \begin{tabular}[c]{@{}c@{}}RA \ (J2000)\\ (deg)\end{tabular} & \begin{tabular}[c]{@{}c@{}}Dec (J2000)\\ (deg)\end{tabular} & ... & \begin{tabular}[c]{@{}c@{}}$T_{\rm eff}$\\ (K)\end{tabular} & \begin{tabular}[c]{@{}c@{}}$\log g$\\ (cm s$^{-2}$)\end{tabular} & $\chi_{\rm red}^{2}$ & \begin{tabular}[c]{@{}c@{}}Luminosity\\ (L$_{\odot}$)\end{tabular} & \begin{tabular}[c]{@{}c@{}} $\pm$\\ (L$_{\odot}$)\end{tabular} & \begin{tabular}[c]{@{}c@{}} Radius\\ (R$_{\odot}$)\end{tabular} & \begin{tabular}[c]{@{}c@{}}$\pm$\\ (R$_{\odot}$)\end{tabular} & \begin{tabular}[c]{@{}c@{}}Mass\\ (M$_{\odot})$\end{tabular} & \begin{tabular}[c]{@{}c@{}}...\\ \end{tabular} & \begin{tabular}[c]{@{}c@{}}Age\\ (Gyr)\end{tabular} & \begin{tabular}[c]{@{}c@{}}...\\ \end{tabular} & \begin{tabular}[c]{@{}c@{}}$S_{{\rm ph},g}$\\ ($\rm ppm$)\end{tabular} &
\begin{tabular}[c]{@{}c@{}}$S_{{\rm ph},r}$\\ ($\rm ppm$)\end{tabular} \\ \hline
ZTFJ000000.13+620605.8 & 0.00056 & 62.10163 & ... & 4200 & 4.5 & 9.82 & 0.215 & 0.014 & 0.87 & 0.025 & 0.8 & ... & 0.03 & ... & 2121 & 1406 \\
ZTFJ000002.20+480720.8 & 0.00918 & 48.12246 & ... & 4300 & 5 & 8.65 & 0.117 & 0.005 & 0.61 & 0.01 & --- & ... & --- & ... & --- & --- \\
ZTFJ000003.23+543605.4 & 0.01347 & 54.60151 & ... & 4200 & 2.5 & 5.15 & 0.224 & 0.018 & 0.90 & 0.03 & 0.81 & ... & 0.028 & ... & 3104 & 2324 \\
ZTFJ000003.76+532917.1 & 0.01568 & 53.48811 & ... & 5400 & 4.5 & 6.01 & 0.694 & 0.04 & 0.95 & 0.02 & 0.91 & ... & 0.037 & ... & 3056 & 2847 \\
ZTFJ000003.34+522031.5 & 0.01809 & 52.34209 & ... & 4700 & 3.5 & 2.95 & 0.278 & 0.032 & 0.795 & 0.043 & 0.78 & ... & 0.05 & ... & 2037 & 1438 \\
... & ... & ... & ... & ... & ... & ... & ... & ... & ... & ... & ... & ... & ... & ... & ... & ...\\
... & ... & ... & ... & ... & ... & ... & ... & ... & ... & ... & ... & ... & ... & ... & ... & ...\\ \hline
\end{tabular}%

}
\label{table2}
\end{table*}

%% file: mnras_deepak_tempelate.bbl
\begin{thebibliography}{}
\expandafter\ifx\csname natexlab\endcsname\relax\def\natexlab#1{#1}\fi
\providecommand{\url}[1]{\href{#1}{#1}}

\bibitem[{{Allard} {et~al.}(2011){Allard}, {Homeier}, \&
  {Freytag}}]{Allard2011}
{Allard}, F., {Homeier}, D., \& {Freytag}, B. 2011, in Astronomical Society of
  the Pacific Conference Series, Vol. 448, 16th Cambridge Workshop on Cool
  Stars, Stellar Systems, and the Sun, ed. C.~{Johns-Krull}, M.~K. {Browning},
  \& A.~A. {West}, 91

\bibitem[{{Bailer-Jones} {et~al.}(2021){Bailer-Jones}, {Rybizki}, {Fouesneau},
  {Demleitner}, \& {Andrae}}]{BailerJones2021}
{Bailer-Jones}, C.~A.~L., {Rybizki}, J., {Fouesneau}, M., {Demleitner}, M., \&
  {Andrae}, R. 2021, AJ, 161, 147

\bibitem[{{Bayo} {et~al.}(2008){Bayo}, {Rodrigo}, {Barrado Y Navascu{\'e}s},
  {Solano}, {Guti{\'e}rrez}, {Morales-Calder{\'o}n}, \& {Allard}}]{Bayo2008}
{Bayo}, A., {Rodrigo}, C., {Barrado Y Navascu{\'e}s}, D., {et~al.} 2008, A\&A,
  492, 277

\bibitem[{{Bopp} \& {Fekel}(1977)}]{Bopp1977}
{Bopp}, B.~W., \& {Fekel}, F., J. 1977, AJ, 82, 490

\bibitem[{{Boro Saikia} {et~al.}(2018){Boro Saikia}, {Marvin}, {Jeffers},
  {Reiners}, {Cameron}, {Marsden}, {Petit}, {Warnecke}, \&
  {Yadav}}]{Saikia2018}
{Boro Saikia}, S., {Marvin}, C.~J., {Jeffers}, S.~V., {et~al.} 2018, A\&A, 616,
  A108

\bibitem[{{Caffau} {et~al.}(2011){Caffau}, {Ludwig}, {Steffen}, {Freytag}, \&
  {Bonifacio}}]{Caffau2011}
{Caffau}, E., {Ludwig}, H.~G., {Steffen}, M., {Freytag}, B., \& {Bonifacio}, P.
  2011, Sol. Phys., 268, 255

\bibitem[{{Casagrande} \& {VandenBerg}(2018)}]{Casagrande2018}
{Casagrande}, L., \& {VandenBerg}, D.~A. 2018, MNRAS, 479, L102

\bibitem[{{Chen} {et~al.}(2020){Chen}, {Wang}, {Deng}, {de Grijs}, {Yang}, \&
  {Tian}}]{Chen2020}
{Chen}, X., {Wang}, S., {Deng}, L., {et~al.} 2020, ApJS, 249, 18

\bibitem[{{Corsaro} {et~al.}(2021){Corsaro}, {Bonanno}, {Mathur},
  {Garc{\'\i}a}, {Santos}, {Breton}, \& {Khalatyan}}]{Corsaro2021}
{Corsaro}, E., {Bonanno}, A., {Mathur}, S., {et~al.} 2021, A\&A, 652, L2

\bibitem[{{Cranmer} \& {Saar}(2011)}]{Cranmer2011}
{Cranmer}, S.~R., \& {Saar}, S.~H. 2011, ApJ, 741, 54

\bibitem[{{Cutri} {et~al.}(2003){Cutri}, {Skrutskie}, {van Dyk}, {Beichman},
  {Carpenter}, {Chester}, {Cambresy}, {Evans}, {Fowler}, {Gizis}, {Howard},
  {Huchra}, {Jarrett}, {Kopan}, {Kirkpatrick}, {Light}, {Marsh}, {McCallon},
  {Schneider}, {Stiening}, {Sykes}, {Weinberg}, {Wheaton}, {Wheelock}, \&
  {Zacarias}}]{Cutri2003}
{Cutri}, R.~M., {Skrutskie}, M.~F., {van Dyk}, S., {et~al.} 2003, VizieR Online
  Data Catalog, II/246

\bibitem[{{Cutri} {et~al.}(2021){Cutri}, {Wright}, {Conrow}, {Fowler},
  {Eisenhardt}, {Grillmair}, {Kirkpatrick}, {Masci}, {McCallon}, {Wheelock},
  {Fajardo-Acosta}, {Yan}, {Benford}, {Harbut}, {Jarrett}, {Lake}, {Leisawitz},
  {Ressler}, {Stanford}, {Tsai}, {Liu}, {Helou}, {Mainzer}, {Gettngs},
  {Gonzalez}, {Hoffman}, {Marsh}, {Padgett}, {Skrutskie}, {Beck}, {Papin}, \&
  {Wittman}}]{Cutri2021}
{Cutri}, R.~M., {Wright}, E.~L., {Conrow}, T., {et~al.} 2021, VizieR Online
  Data Catalog, II/328

\bibitem[{{Davenport} \& {Covey}(2018)}]{Davenport2018}
{Davenport}, J. R.~A., \& {Covey}, K.~R. 2018, ApJ, 868, 151

\bibitem[{{de Grijs} \& {Kamath}(2021)}]{deGrijs2021}
{de Grijs}, R., \& {Kamath}, D. 2021, Universe, 7, 440

\bibitem[{{Gaia Collaboration} {et~al.}(2021){Gaia Collaboration}, {Brown},
  {Vallenari}, {Prusti}, {de Bruijne}, {Babusiaux}, {Biermann}, {Creevey},
  {Evans}, {Eyer}, {Hutton}, {Jansen}, {Jordi}, {Klioner}, {Lammers},
  {Lindegren}, {Luri}, {Mignard}, {Panem}, {Pourbaix}, {Randich}, {Sartoretti},
  {Soubiran}, {Walton}, {Arenou}, {Bailer-Jones}, {Bastian}, {Cropper},
  {Drimmel}, {Katz}, {Lattanzi}, {van Leeuwen}, {Bakker}, {Cacciari},
  {Casta{\~n}eda}, {De Angeli}, {Ducourant}, {Fabricius}, {Fouesneau},
  {Fr{\'e}mat}, {Guerra}, {Guerrier}, {Guiraud}, {Jean-Antoine Piccolo},
  {Masana}, {Messineo}, {Mowlavi}, {Nicolas}, {Nienartowicz}, {Pailler},
  {Panuzzo}, {Riclet}, {Roux}, {Seabroke}, {Sordo}, {Tanga}, {Th{\'e}venin},
  {Gracia-Abril}, {Portell}, {Teyssier}, {Altmann}, {Andrae}, {Bellas-Velidis},
  {Benson}, {Berthier}, {Blomme}, {Brugaletta}, {Burgess}, {Busso}, {Carry},
  {Cellino}, {Cheek}, {Clementini}, {Damerdji}, {Davidson}, {Delchambre},
  {Dell'Oro}, {Fern{\'a}ndez-Hern{\'a}ndez}, {Galluccio}, {Garc{\'\i}a-Lario},
  {Garcia-Reinaldos}, {Gonz{\'a}lez-N{\'u}{\~n}ez}, {Gosset}, {Haigron},
  {Halbwachs}, {Hambly}, {Harrison}, {Hatzidimitriou}, {Heiter},
  {Hern{\'a}ndez}, {Hestroffer}, {Hodgkin}, {Holl}, {Jan{\ss}en}, {Jevardat de
  Fombelle}, {Jordan}, {Krone-Martins}, {Lanzafame}, {L{\"o}ffler}, {Lorca},
  {Manteiga}, {Marchal}, {Marrese}, {Moitinho}, {Mora}, {Muinonen}, {Osborne},
  {Pancino}, {Pauwels}, {Petit}, {Recio-Blanco}, {Richards}, {Riello},
  {Rimoldini}, {Robin}, {Roegiers}, {Rybizki}, {Sarro}, {Siopis}, {Smith},
  {Sozzetti}, {Ulla}, {Utrilla}, {van Leeuwen}, {van Reeven}, {Abbas}, {Abreu
  Aramburu}, {Accart}, {Aerts}, {Aguado}, {Ajaj}, {Altavilla}, {{\'A}lvarez},
  {{\'A}lvarez Cid-Fuentes}, {Alves}, {Anderson}, {Anglada Varela}, {Antoja},
  {Audard}, {Baines}, {Baker}, {Balaguer-N{\'u}{\~n}ez}, {Balbinot}, {Balog},
  {Barache}, {Barbato}, {Barros}, {Barstow}, {Bartolom{\'e}}, {Bassilana},
  {Bauchet}, {Baudesson-Stella}, {Becciani}, {Bellazzini}, {Bernet}, {Bertone},
  {Bianchi}, {Blanco-Cuaresma}, {Boch}, {Bombrun}, {Bossini}, {Bouquillon},
  {Bragaglia}, {Bramante}, {Breedt}, {Bressan}, {Brouillet}, {Bucciarelli},
  {Burlacu}, {Busonero}, {Butkevich}, {Buzzi}, {Caffau}, {Cancelliere},
  {C{\'a}novas}, {Cantat-Gaudin}, {Carballo}, {Carlucci}, {Carnerero},
  {Carrasco}, {Casamiquela}, {Castellani}, {Castro-Ginard}, {Castro Sampol},
  {Chaoul}, {Charlot}, {Chemin}, {Chiavassa}, {Cioni}, {Comoretto}, {Cooper},
  {Cornez}, {Cowell}, {Crifo}, {Crosta}, {Crowley}, {Dafonte}, {Dapergolas},
  {David}, {David}, {de Laverny}, {De Luise}, {De March}, {De Ridder}, {de
  Souza}, {de Teodoro}, {de Torres}, {del Peloso}, {del Pozo}, {Delbo},
  {Delgado}, {Delgado}, {Delisle}, {Di Matteo}, {Diakite}, {Diener},
  {Distefano}, {Dolding}, {Eappachen}, {Edvardsson}, {Enke}, {Esquej}, {Fabre},
  {Fabrizio}, {Faigler}, {Fedorets}, {Fernique}, {Fienga}, {Figueras},
  {Fouron}, {Fragkoudi}, {Fraile}, {Franke}, {Gai}, {Garabato},
  {Garcia-Gutierrez}, {Garc{\'\i}a-Torres}, {Garofalo}, {Gavras}, {Gerlach},
  {Geyer}, {Giacobbe}, {Gilmore}, {Girona}, {Giuffrida}, {Gomel}, {Gomez},
  {Gonzalez-Santamaria}, {Gonz{\'a}lez-Vidal}, {Granvik},
  {Guti{\'e}rrez-S{\'a}nchez}, {Guy}, {Hauser}, {Haywood}, {Helmi}, {Hidalgo},
  {Hilger}, {H{\l}adczuk}, {Hobbs}, {Holland}, {Huckle}, {Jasniewicz},
  {Jonker}, {Juaristi Campillo}, {Julbe}, {Karbevska}, {Kervella}, {Khanna},
  {Kochoska}, {Kontizas}, {Kordopatis}, {Korn}, {Kostrzewa-Rutkowska},
  {Kruszy{\'n}ska}, {Lambert}, {Lanza}, {Lasne}, {Le Campion}, {Le Fustec},
  {Lebreton}, {Lebzelter}, {Leccia}, {Leclerc}, {Lecoeur-Taibi}, {Liao},
  {Licata}, {Lindstr{\o}m}, {Lister}, {Livanou}, {Lobel}, {Madrero Pardo},
  {Managau}, {Mann}, {Marchant}, {Marconi}, {Marcos Santos}, {Marinoni},
  {Marocco}, {Marshall}, {Martin Polo}, {Mart{\'\i}n-Fleitas}, {Masip},
  {Massari}, {Mastrobuono-Battisti}, {Mazeh}, {McMillan}, {Messina},
  {Michalik}, {Millar}, {Mints}, {Molina}, {Molinaro}, {Moln{\'a}r},
  {Montegriffo}, {Mor}, {Morbidelli}, {Morel}, {Morris}, {Mulone}, {Munoz},
  {Muraveva}, {Murphy}, {Musella}, {Noval}, {Ord{\'e}novic}, {Orr{\`u}},
  {Osinde}, {Pagani}, {Pagano}, {Palaversa}, {Palicio}, {Panahi}, {Pawlak},
  {Pe{\~n}alosa Esteller}, {Penttil{\"a}}, {Piersimoni}, {Pineau}, {Plachy},
  {Plum}, {Poggio}, {Poretti}, {Poujoulet}, {Pr{\v{s}}a}, {Pulone}, {Racero},
  {Ragaini}, {Rainer}, {Raiteri}, {Rambaux}, {Ramos}, {Ramos-Lerate}, {Re
  Fiorentin}, {Regibo}, {Reyl{\'e}}, {Ripepi}, {Riva}, {Rixon}, {Robichon},
  {Robin}, {Roelens}, {Rohrbasser}, {Romero-G{\'o}mez}, {Rowell}, {Royer},
  {Rybicki}, {Sadowski}, {Sagrist{\`a} Sell{\'e}s}, {Sahlmann}, {Salgado},
  {Salguero}, {Samaras}, {Sanchez Gimenez}, {Sanna}, {Santove{\~n}a},
  {Sarasso}, {Schultheis}, {Sciacca}, {Segol}, {Segovia}, {S{\'e}gransan},
  {Semeux}, {Shahaf}, {Siddiqui}, {Siebert}, {Siltala}, {Slezak}, {Smart},
  {Solano}, {Solitro}, {Souami}, {Souchay}, {Spagna}, {Spoto}, {Steele},
  {Steidelm{\"u}ller}, {Stephenson}, {S{\"u}veges}, {Szabados}, {Szegedi-Elek},
  {Taris}, {Tauran}, {Taylor}, {Teixeira}, {Thuillot}, {Tonello}, {Torra},
  {Torra}, {Turon}, {Unger}, {Vaillant}, {van Dillen}, {Vanel}, {Vecchiato},
  {Viala}, {Vicente}, {Voutsinas}, {Weiler}, {Wevers}, {Wyrzykowski}, {Yoldas},
  {Yvard}, {Zhao}, {Zorec}, {Zucker}, {Zurbach}, \&
  {Zwitter}}]{GaiaCollaboration2021}
{Gaia Collaboration}, {Brown}, A.~G.~A., {Vallenari}, A., {et~al.} 2021, A\&A,
  649, A1

\bibitem[{{Gordon} {et~al.}(2021){Gordon}, {Davenport}, {Angus},
  {Foreman-Mackey}, {Agol}, {Covey}, {Ag{\"u}eros}, \& {Kipping}}]{Gordon2021}
{Gordon}, T.~A., {Davenport}, J. R.~A., {Angus}, R., {et~al.} 2021, ApJ, 913,
  70

\bibitem[{{Green} {et~al.}(2019){Green}, {Schlafly}, {Zucker}, {Speagle}, \&
  {Finkbeiner}}]{Green2019}
{Green}, G.~M., {Schlafly}, E., {Zucker}, C., {Speagle}, J.~S., \&
  {Finkbeiner}, D. 2019, ApJ, 887, 93

\bibitem[{{Hall}(2008)}]{Hall2008}
{Hall}, J.~C. 2008, Living Reviews in Solar Physics, 5, 2

\bibitem[{{Kubiak} {et~al.}(2021){Kubiak}, {Mu{\v{z}}i{\'c}}, {Sousa},
  {Almendros-Abad}, {K{\"o}hler}, \& {Scholz}}]{Kubiak2021}
{Kubiak}, K., {Mu{\v{z}}i{\'c}}, K., {Sousa}, I., {et~al.} 2021, A\&A, 650, A48

\bibitem[{{Lanzafame} {et~al.}(2018){Lanzafame}, {Distefano}, {Messina},
  {Pagano}, {Lanza}, {Eyer}, {Guy}, {Rimoldini}, {Lecoeur-Taibi}, {Holl},
  {Audard}, {de Fombelle}, {Nienartowicz}, {Marchal}, \&
  {Mowlavi}}]{Lanzafame2018}
{Lanzafame}, A.~C., {Distefano}, E., {Messina}, S., {et~al.} 2018, A\&A, 616,
  A16

\bibitem[{{Lomb}(1976)}]{Lomb1976}
{Lomb}, N.~R. 1976, Ap\&SS, 39, 447

\bibitem[{{Mainzer} {et~al.}(2011){Mainzer}, {Grav}, {Bauer}, {Masiero},
  {McMillan}, {Cutri}, {Walker}, {Wright}, {Eisenhardt}, {Tholen}, {Spahr},
  {Jedicke}, {Denneau}, {DeBaun}, {Elsbury}, {Gautier}, {Gomillion}, {Hand},
  {Mo}, {Watkins}, {Wilkins}, {Bryngelson}, {Del Pino Molina}, {Desai},
  {G{\'o}mez Camus}, {Hidalgo}, {Konstantopoulos}, {Larsen}, {Maleszewski},
  {Malkan}, {Mauduit}, {Mullan}, {Olszewski}, {Pforr}, {Saro}, {Scotti}, \&
  {Wasserman}}]{Mainzer2011}
{Mainzer}, A., {Grav}, T., {Bauer}, J., {et~al.} 2011, ApJ, 743, 156

\bibitem[{{Masci} {et~al.}(2019){Masci}, {Laher}, {Rusholme}, {Shupe}, {Groom},
  {Surace}, {Jackson}, {Monkewitz}, {Beck}, {Flynn}, {Terek}, {Landry},
  {Hacopians}, {Desai}, {Howell}, {Brooke}, {Imel}, {Wachter}, {Ye}, {Lin},
  {Cenko}, {Cunningham}, {Rebbapragada}, {Bue}, {Miller}, {Mahabal}, {Bellm},
  {Patterson}, {Juri{\'c}}, {Golkhou}, {Ofek}, {Walters}, {Graham}, {Kasliwal},
  {Dekany}, {Kupfer}, {Burdge}, {Cannella}, {Barlow}, {Van Sistine}, {Giomi},
  {Fremling}, {Blagorodnova}, {Levitan}, {Riddle}, {Smith}, {Helou}, {Prince},
  \& {Kulkarni}}]{Masci2019}
{Masci}, F.~J., {Laher}, R.~R., {Rusholme}, B., {et~al.} 2019, Publ. Astron.
  Soc. Pac., 131, 018003

\bibitem[{{Mathur} {et~al.}(2014){Mathur}, {Salabert}, {Garc{\'\i}a}, \&
  {Ceillier}}]{Mathur2014b}
{Mathur}, S., {Salabert}, D., {Garc{\'\i}a}, R.~A., \& {Ceillier}, T. 2014,
  Journal of Space Weather and Space Climate, 4, A15

\bibitem[{{McQuillan} {et~al.}(2014){McQuillan}, {Mazeh}, \&
  {Aigrain}}]{Mcquillan2014}
{McQuillan}, A., {Mazeh}, T., \& {Aigrain}, S. 2014, ApJS, 211, 24

\bibitem[{{Noyes} {et~al.}(1984){Noyes}, {Hartmann}, {Baliunas}, {Duncan}, \&
  {Vaughan}}]{Noyes1984}
{Noyes}, R.~W., {Hartmann}, L.~W., {Baliunas}, S.~L., {Duncan}, D.~K., \&
  {Vaughan}, A.~H. 1984, ApJ, 279, 763

\bibitem[{{Parker}(1955)}]{Parker1955}
{Parker}, E.~N. 1955, ApJ, 122, 293

\bibitem[{{Pecaut} \& {Mamajek}(2013)}]{Pecaut2013}
{Pecaut}, M.~J., \& {Mamajek}, E.~E. 2013, ApJS, 208, 9

\bibitem[{{Pizzolato} {et~al.}(2003){Pizzolato}, {Maggio}, {Micela},
  {Sciortino}, \& {Ventura}}]{Pizzolato2003}
{Pizzolato}, N., {Maggio}, A., {Micela}, G., {Sciortino}, S., \& {Ventura}, P.
  2003, A\&A, 397, 147

\bibitem[{{Reiners} {et~al.}(2014){Reiners}, {Sch{\"u}ssler}, \&
  {Passegger}}]{Reiners2014}
{Reiners}, A., {Sch{\"u}ssler}, M., \& {Passegger}, V.~M. 2014, ApJ, 794, 144

\bibitem[{{Reinhold} \& {Hekker}(2020)}]{Reinhold2020}
{Reinhold}, T., \& {Hekker}, S. 2020, A\&A, 635, A43

\bibitem[{{Reinhold} {et~al.}(2013){Reinhold}, {Reiners}, \&
  {Basri}}]{Reinhold2013}
{Reinhold}, T., {Reiners}, A., \& {Basri}, G. 2013, A\&A, 560, A4

\bibitem[{{Reinhold} {et~al.}(2020){Reinhold}, {Shapiro}, {Solanki}, {Montet},
  {Krivova}, {Cameron}, \& {Amazo-G{\'o}mez}}]{Reinhold2020_Sun}
{Reinhold}, T., {Shapiro}, A.~I., {Solanki}, S.~K., {et~al.} 2020, Science,
  368, 518

\bibitem[{{Salabert} {et~al.}(2016){Salabert}, {Garc{\'\i}a}, {Beck},
  {Egeland}, {Pall{\'e}}, {Mathur}, {Metcalfe}, {do Nascimento}, {Ceillier},
  {Andersen}, \& {Trivi{\~n}o Hage}}]{Salabert2016}
{Salabert}, D., {Garc{\'\i}a}, R.~A., {Beck}, P.~G., {et~al.} 2016, A\&A, 596,
  A31

\bibitem[{{Scargle}(1982)}]{Scargle1982}
{Scargle}, J.~D. 1982, ApJ, 263, 835

\bibitem[{{See} {et~al.}(2021){See}, {Roquette}, {Amard}, \& {Matt}}]{See2021}
{See}, V., {Roquette}, J., {Amard}, L., \& {Matt}, S.~P. 2021, \apj, 912, 127

\bibitem[{{Skumanich}(1972)}]{Skumanich1972}
{Skumanich}, A. 1972, ApJ, 171, 565

\bibitem[{{Somers} {et~al.}(2020){Somers}, {Cao}, \&
  {Pinsonneault}}]{Somers2020}
{Somers}, G., {Cao}, L., \& {Pinsonneault}, M.~H. 2020, ApJ, 891, 29

\bibitem[{{Stassun} \& {Torres}(2021)}]{Sassun2021}
{Stassun}, K.~G., \& {Torres}, G. 2021, ApJ, 907, L33

\bibitem[{{Strassmeier}(2009)}]{Strassmeier2009}
{Strassmeier}, K.~G. 2009, A\&ARv, 17, 251

\bibitem[{{van Saders} \& {Pinsonneault}(2012)}]{vanSaders2012}
{van Saders}, J.~L., \& {Pinsonneault}, M.~H. 2012, ApJ, 746, 16

\bibitem[{{Vaughan} \& {Preston}(1980)}]{Vaughan1980}
{Vaughan}, A.~H., \& {Preston}, G.~W. 1980, Publ. Astron. Soc. Pac., 92, 385

\bibitem[{{Wang} \& {Chen}(2019)}]{Wang2019}
{Wang}, S., \& {Chen}, X. 2019, ApJ, 877, 116

\bibitem[{{Wilson}(1968)}]{Wilson1968}
{Wilson}, O.~C. 1968, ApJ, 153, 221

\bibitem[{{Wright} {et~al.}(2010){Wright}, {Eisenhardt}, {Mainzer}, {Ressler},
  {Cutri}, {Jarrett}, {Kirkpatrick}, {Padgett}, {McMillan}, {Skrutskie},
  {Stanford}, {Cohen}, {Walker}, {Mather}, {Leisawitz}, {Gautier}, {McLean},
  {Benford}, {Lonsdale}, {Blain}, {Mendez}, {Irace}, {Duval}, {Liu}, {Royer},
  {Heinrichsen}, {Howard}, {Shannon}, {Kendall}, {Walsh}, {Larsen}, {Cardon},
  {Schick}, {Schwalm}, {Abid}, {Fabinsky}, {Naes}, \& {Tsai}}]{Wright2010}
{Wright}, E.~L., {Eisenhardt}, P. R.~M., {Mainzer}, A.~K., {et~al.} 2010, AJ,
  140, 1868

\bibitem[{{Zhang} {et~al.}(2019){Zhang}, {Zhao}, {Oswalt}, {Fang}, {Zhao},
  {Liang}, {Ye}, \& {Zhong}}]{Zhang2019}
{Zhang}, J., {Zhao}, J., {Oswalt}, T.~D., {et~al.} 2019, ApJ, 887, 84

\end{thebibliography}
